\documentclass[%
aps,
superscriptaddress,nofootinbib,%floatfix
reprint,
]{revtex4-2}
\usepackage[mathscr]{euscript}
\usepackage{bm}% bold math
\usepackage{graphicx}% Include figure files
\usepackage{subfigure}
\usepackage{overpic}
\usepackage{multirow}
\usepackage{booktabs}
\usepackage{array}
\usepackage{floatrow}
\usepackage{float} %for fixing the figure
\usepackage{amsmath,amssymb,amsthm}
\usepackage{cases}
\usepackage[colorlinks=true,linkcolor=red]{hyperref}
\newfloatcommand{capbtabbox}{table}[][\FBwidth]
\newcommand{\bea}{\begin{eqnarray}}
\newcommand{\eea}{\end{eqnarray}}
\newcommand{\beq}{\begin{equation}}
\newcommand{\eeq}{\end{equation}}

\def\/{\over}

\begin{document}
%%%%%%%%%%%%%%%%%%%%%%%%%%%%%%%%%%%%%%%%%%%%%%%%%%%%%%%%%%%%%%%%%%%%%%%%%%%%%%%%
	
%%%%%%%%%%%%%%%%%%%%%%%%%%%%%%%%%%%% title %%%%%%%%%%%%%%%%%%%%%%%%%%%%%%%%%%%%%
\title{Primordial Gravitational Waves in Parity-violating Symmetric Teleparallel Gravity}

%%%%%%%%%%%%%%%%%%%%%%%%%%%%%%%%%%%% author %%%%%%%%%%%%%%%%%%%%%%%%%%%%%%%%%%%%

\author{ Rongrong Zhai}
\email{rrzhai@xztu.edu.cn}
\affiliation{Department of Physics, Xinzhou Normal University, Xinzhou 034000, Shanxi, China}

\author{Chengjie Fu}
\email[Corresponding author:~]{fucj@ahnu.edu.cn}
\affiliation{Department of Physics, Anhui Normal University, Wuhu, Anhui 241002, China}

\author{Xiangyun Fu}
\email{xyfu@hnust.edu.cn}
\affiliation{School of Physics and Electronic Science, Hunan University of Science and Technology,
 Xiangtan, Hunan 411021, China}

\author{Puxun Wu}
\email{pxwu@hunnu.edu.cn}
\affiliation{Department of Physics and Synergetic Innovation Center for Quantum Effects and Applications, Hunan Normal University, Changsha, Hunan 410081, China}
\affiliation{Institute of Interdisciplinary Studies, Hunan Normal University, Changsha, Hunan 410081, China}

\author{Hongwei Yu}
\email{hwyu@hunnu.edu.cn}
\affiliation{Department of Physics and Synergetic Innovation Center for Quantum Effects and Applications, Hunan Normal University, Changsha, Hunan 410081, China}
\affiliation{Institute of Interdisciplinary Studies, Hunan Normal University, Changsha, Hunan 410081, China}
%%%%%%%%%%%%%%%%%%%%%%%%%%%%%%%%%%%% author %%%%%%%%%%%%%%%%%%%%%%%%%%%%%%%%%%%%

%%%%%%%%%%%%%%%%%%%%%%%%%%%%%%%%% abstract %%%%%%%%%%%%%%%%%%%%%%%%%%%%%%%%%%%%%
\begin{abstract}
In this paper, we investigate the inflationary phenomenology of parity-violating (PV) extensions of symmetric teleparallel gravity by applying this PV gravity theory to axion inflation. The presence of PV terms induces velocity birefringence in the tensor perturbations. During inflation, when the inflaton rapidly traverses the cliff-like region in its potential, the tensor modes at specific scales for one of the two circular polarization states undergo significant amplification due to tachyonic instability. 
Consequently, the resulting primordial gravitational waves (GWs), characterized by a one-handed polarization and a multi-peak structure in their energy spectrum, exhibit a significant amplitude potentially detectable by LISA and Taiji, and their chirality could be determined by the LISA-Taiji network. The detection of such a chiral GW signal provides an opportunity to probe inflation and PV gravity theory. Moreover, we perform the Fisher matrix analysis to forecast the constraints on the model parameters with the LISA-Taiji network.
%Furthermore, by comparing the predicted energy spectra with the sensitivity curves of these detectors, we quantitatively determine the minimum coupling threshold required for a reliable detection. This result indicates that the detectability and chirality of the primordial GW background are highly sensitive to the coupling constant. Therefore, if such signals are observed in the future, stringent constraints can be imposed on the parameters of the PV sector in the STG theory.

\end{abstract}

\maketitle

%%%%%%%%%%%%%%%%%%%%%%%%%%%%%%%%%%%%%%%%%%%%%%%%%%%%%%%%%%%%%%%%%%%%%%%%%%%%%%%%
\section{Introduction}
\label{sec_in}

The detection of gravitational waves (GWs) by the LIGO-Virgo collaboration \cite{LIGOScientific:2016sjg,*LIGOScientific:2016aoc,*LIGOScientific:2017vox,*LIGOScientific:2017vwq,*LIGOScientific:2017ycc,*LIGOScientific:2018mvr,*LIGOScientific:2020aai} has opened a new observational window for probing the nature of gravity in the strong-field and nonlinear regime.
These GW signals from astrophysical sources not only confirm key predictions of general relativity (GR) but also spark renewed interest in the GW background originating from the early universe. In particular, primordial GWs, also known as tensor perturbations, generated during inflation carry valuable information about the physics of the early universe and the fundamental nature of gravity \cite{Guzzetti:2016mkm,Bartolo:2016ami,LISACosmologyWorkingGroup:2022jok,Cai:2017cbj,Caprini:2018mtu,Kuroyanagi:2018csn}. Measurements of the cosmic microwave background (CMB) constrain the amplitude of primordial tensor perturbations at $f\sim10^{-17}\rm{Hz}$. Specifically, current CMB observations provide an upper bound on the tensor-to-scalar ratio $r$ at the pivot scale $k_\ast=0.05\rm{Mpc}^{-1}$, with $r< 0.036$ at 95\% confidence \cite{BICEP:2021xfz}. Apart from observations at very large CMB scales, GW interferometers offer promising opportunities to probe the primordial GWs at much smaller scales. Future experiments such as LISA \cite{LISA:2017pwj,Barausse:2020rsu} and Taiji \cite{Hu:2017mde,Ruan:2018tsw} are expected to be sensitive to signals peaking in the frequency range $f\sim \mathcal{O}(10^{-4}-10^{-1}){\rm Hz}$, which corresponds to tensor perturbations with $k\sim \mathcal{O}(10^{11}-10^{14}){\rm Mpc}^{-1}$. However, for the simplest realization of inflation, the resulting primordial GWs typically exhibit an almost scale-invariant spectrum with an amplitude that is too small to be detected by current or near-future GW interferometers.

In the literature, there has been extensive exploration of mechanisms that can amplify the primordial tensor spectrum at small scales. For instance, such amplification may result from the breaking of space-reparametrization symmetry during inflation \cite{Cannone:2014uqa}, or from a non-attractor phase in the context of generalized G-inflation \cite{Mylova:2018yap,Ozsoy:2019slf}. Moreover, a prominent class of scenarios is built upon the framework of parity-violating (PV) gravity theories, such as the well-known Chern-Simons modified gravity \cite{Lue:1998mq,Jackiw:2003pm,Alexander:2009tp}, ghost-free PV scalar-tensor theory \cite{Crisostomi:2017ugk,Nishizawa:2018srh,Gao:2019liu}, and PV extensions of teleparallel gravity \cite{Li:2020xjt,Li:2022mti} and symmetric teleparallel gravity (STG) \cite{Conroy:2019ibo,Li:2021mdp,Li:2022vtn}. A characteristic feature of these theories is that the left- and right-handed polarization modes of GWs obey distinct equations of motion, leading to rich phenomenological consequences \cite{Alexander:2004us,Satoh:2007gn,Saito:2007kt,Gluscevic:2010vv,Wang:2012fi,Alexander:2016hxk,Bartolo:2017szm,Qiao:2019hkz,Wang:2020cub,Okounkova:2021xjv,Wu:2021ndf,Gong:2021jgg,Odintsov:2022hxu,Zhu:2022uoq,Zhu:2022dfq}.
Based on Chern-Simons modified gravity, it has been shown that small-scale tensor perturbations can be significantly enhanced through parametric resonance during inflation \cite{Fu:2020tlw,Peng:2022ttg}. Alternatively, by applying Nieh-Yan modified Teleparallel Gravity (NYmTG) to inflationary scenarios, Refs. \cite{Cai:2021uup,Fu:2023aab} have shown that one of the two circular polarization states of primordial GWs can be significantly amplified at small scales due to the tachyonic instability. 

Motivated by previous developments, in this paper we investigate the amplification of small-scale primordial GWs within the theoretical framework of PV extensions of STG, yielding a GW signal that may be detectable by next-generation interferometers such as LISA and Taiji. The STG theories attribute gravity to the nonmetricity of a flat, torsion-free connection. 
Within this framework, one may have a model dynamically equivalent to GR, known as the Symmetric Teleparallel Equivalent of General Relativity (STEGR), which produces the same field equations as GR \cite{Nester:1998mp,BeltranJimenez:2019esp,Capozziello:2022zzh}. The PV extensions of STG are constructed by extending STEGR through the inclusion of PV interactions between the scalar field and parity-odd terms that are quadratic in the nonmetricity tensor \cite{Conroy:2019ibo,Li:2022vtn,Li:2021mdp}. These PV terms induce a difference between the propagating velocities of the left- and right-handed polarized states of GWs, a phenomenon known as velocity birefringence. We expect that this phenomenon, when combined with the axion inflation, can lead to a primordial GW signal that is potentially detectable by future GW observations.

The organization of this paper is as follows.
Section~\ref{sec2} provides a brief review of the STG framework with PV extensions, derives the corresponding field equations in a spatially flat Friedmann-Robertson-Walker (FRW) background, and briefly analyzes the behavior of tensor perturbations during inflation.
In Sec.~\ref{sec3}, we apply this model to axion inflation and demonstrate that the PV terms can generate a chiral GW signal potentially detectable by future space-based interferometers such as LISA and Taiji.
In Sec.~\ref{sec4}, we perform parameter estimation with the LISA-Taiji network.
Our conclusions are presented in Sec.~\ref{conclusion}. 
Throughout this work, we adopt the natural units with $c=\hbar=1$, and set the reduced Planck mass as $M_{\rm Pl}=1/\sqrt{8\pi G}=1$. Greek indices $\mu,\nu,\rho,\dots=0,1,2,3$ denote spacetime components, while Latin indices $i,j,k,\dots=1,2,3$ represent spatial components.

%%%%%%%%%%%%%%%%%%%%%%%%%%%%%%%%%%%%%%%%%%%%%%%%%%%%%%%%%%%%%%%%%%%%%%%%%%%%%%%%

\section{Tensor Perturbations in the Parity-violating Extensions of Symmetric Teleparallel Gravity}
\label{sec2}

The STG theory is formulated using a metric $g_{\mu\nu}$ and an affine connection ${\Gamma^\lambda}_{\mu\nu}$ that is curvature free and torsionless, and can be expressed as follows:
\begin{equation}
{\Gamma^\lambda}_{\mu\nu}=\frac{\partial x^\lambda}{\partial y^\alpha}\partial_\mu\partial_\nu y^\alpha~,
\end{equation}
where the four functions ${y^\alpha}$ constitute a special coordinate system in which the affine connection vanishes, i.e., ${\Gamma^\rho}_{\mu\nu}=0$. The connection possesses a nonzero nonmetricity tensor, defined as
\begin{equation}
	Q_{\alpha\mu\nu}\equiv \nabla_\alpha g_{\mu\nu}=\partial_\alpha g_{\mu\nu}-{\Gamma^\lambda}_{\alpha\mu}g_{\lambda\nu}-{\Gamma^\lambda}_{\alpha\nu}g_{\mu\lambda}, 
\end{equation}
which encodes the gravitational interaction in the framework of STG theory. Based on the nonmetricity tensor, one can construct an equivalent representation of GR, known as STEGR, whose action is given by
\begin{align}\label{action_STEGR}
	S_{\rm g}=&\frac{1}{2}\int {\rm d}^4x \sqrt{-g}\mathbb{Q} \nonumber \\
    \equiv& \frac{1}{2}\int {\rm d}^4x \sqrt{-g} \left( \frac{1}{4}Q_{\alpha\mu\nu}Q^{\alpha\mu\nu}-\frac{1}{2}Q_{\alpha\mu\nu}Q^{\mu\nu\alpha} \right. \nonumber \\
    &\left. -\frac{1}{4}Q_{\alpha}Q^{\alpha}+\frac{1}{2}Q_{\alpha}\bar{Q}^{\alpha}\right),
\end{align}
where $\mathbb{Q}$ denotes the nonmetricity scalar, $Q_\alpha=Q_{\alpha\mu\nu}g^{\mu\nu}$ and $\bar{Q}_\alpha=Q_{\rho\sigma\alpha}g^{\rho\sigma}$ are two nonmetricity vectors.
Note that the STEGR action is dynamically equivalent to the Einstein-Hilbert action of GR, differing only by a boundary term:
\begin{equation}
	S_{\rm g}=\frac{1}{2}\int {\rm d}^4x \sqrt{-g} \left[-R-\nabla_{\alpha}(Q^\alpha-\bar{Q}^\alpha)\right],
\end{equation}
where both the curvature scalar $R$ and the covariant derivative $\nabla_{\alpha}$ are associated with the Levi-Civita connection. 

PV extensions of STG can be constructed by introducing several interactions between the scalar field and the parity-odd terms~\cite{Li:2022vtn}:
\begin{align}\label{M_i}
M_1&=\varepsilon^{\mu\nu\rho\sigma}\,Q_{\mu\nu\alpha}\,{Q_{\rho\sigma}}^\alpha\,\nabla_\beta \phi\,\nabla^\beta \phi, \nonumber \\
M_2&=\varepsilon^{\mu\nu\rho\sigma}\,Q_{\mu\nu\alpha}\,{Q_{\rho\sigma}}^\beta\,\nabla^\alpha \phi\,\nabla_\beta \phi, \nonumber\\
M_3&=\varepsilon^{\mu\nu\rho\sigma}\,Q_{\mu\nu\alpha}\,{Q_{\rho}}^{\alpha\beta}\,\nabla_\sigma \phi\,\nabla_\beta \phi, \nonumber\\
M_4&=\varepsilon^{\mu\nu\rho\sigma}\,Q_{\mu\nu\alpha}\,{Q^{\alpha\beta}}_\rho\,\nabla_\sigma \phi\,\nabla_\beta \phi, \nonumber\\
M_5&=\varepsilon^{\mu\nu\rho\sigma}\,Q_{\mu\nu\alpha}\,{Q^{\beta\alpha}}_\rho\,\nabla_\sigma \phi\,\nabla_\beta \phi, \nonumber\\ M_6&=\varepsilon^{\mu\nu\rho\sigma}\,Q_{\mu\nu\alpha}\,Q_{\rho}\,\nabla^\alpha\phi\,\nabla_\sigma \phi, \nonumber \\
M_7&=\varepsilon^{\mu\nu\rho\sigma}\,Q_{\mu\nu\alpha}\,\bar{Q}_{\rho}\,\nabla^\alpha\phi\,\nabla_\sigma \phi,
\end{align}
which ensure that the equations of motion of the metric and the scalar field remain second order. Here, $\varepsilon^{\mu\nu\rho\sigma}=\epsilon^{\mu\nu\rho\sigma}/\sqrt{-g}$ is the Levi-Civita tensor with $\epsilon^{\mu\nu\rho\sigma}$ being the totally antisymmetric symbol. The action of parity violation can then be written as
\begin{equation}\label{S_PV}
S_{\rm PV}=\int {\rm d}^4x\sqrt{-g}\sum_{i=1}^7 c_i(\phi,\nabla^\mu \phi \nabla_\mu \phi) M_i,
\end{equation}
where the coupling coefficients $c_i$ may depend on both the scalar field and its first derivatives. The full action for the PV model in the framework of STG is therefore given by
\begin{align}\label{full_action}
S=& S_{\rm g}+S_{\rm PV}+S_\phi \nonumber \\
=&\int {\rm d}^4x\sqrt{-g}\left[\frac{\mathbb{Q}}{2}+\sum_{i=1}^7 c_i(\phi,\nabla^\mu \phi \nabla_\mu \phi) M_i \right. \nonumber  \\ 
& \left. +\frac{1}{2} g^{\mu\nu}\partial_\mu\phi\partial_\nu\phi-V(\phi)\right].
\end{align}

In a spatially flat FRW universe, the PV terms appearing in the action \eqref{full_action} do not affect the background evolution. Consequently, the background equations remain identical to those in GR. The quadratic action for linear scalar perturbations, described by the gauge-invariant variable associated with $\phi$, is identical to that in GR with a minimally coupled scalar field $\phi$. This implies that if the $\phi$ field serves as the inflaton, the primordial scalar perturbations predicted by the model \eqref{full_action} are the same as those in the standard single-field slow-roll inflation. Moreover, the linear vector perturbations in this model are dynamical and exhibit a ghost instability. However, the vector perturbations become nondynamical and the model remains ghostfree if the special condition $2c_1 + 2c_2 - c_4 - c_5 - c_7 = 0$ is imposed. In this paper, we focus on the linear tensor perturbations $h_{ij}$, and thus consider the following perturbed metric, 
\begin{equation}
ds^2= {\rm d}t^2-a^2\left(\delta_{ij}+h_{ij}\right){\rm d}x^i {\rm d}x^j.
\end{equation}
The background equations are given by
\begin{equation}\label{BG_Eq}
3H^2=\frac{1}{2}\dot\phi^2 + V(\phi),\quad \ddot\phi+3H\dot\phi+\frac{{\rm d}V}{{\rm d}\phi}=0.
\end{equation}
Then we decompose the tensor perturbations in terms of the circular polarization bases as
\begin{equation}
h_{ij}=\sum_{A=L,R}\int \frac{{\rm d}^3\vec{k}}{(2\pi)^{3/2}}e^{i\vec{k}\cdot\vec{x}}\,h_A(t,\vec{k})\,e^A_{ij}(\vec{k}),
\end{equation}
where $A=L,R$ denote the left- and right-handed polarization states, respectively. The polarization tensors $e^A_{ij}$ satisfy the relations $e^A_{ij}e^{B\ast}_{ij}=\delta^{AB}$ and $i\epsilon_{ijk}k_i e^A_{jl}=\lambda_A k e^A_{kl}$, with $\lambda_R=1$ and $\lambda_L=-1$. From the action \eqref{full_action}, the mode functions $h_A(t,\vec{k})$ obey the following equation of motion,
\begin{align}\label{EoM_h_A}
\ddot h_A + 3H \dot h_A + \frac{k}{a} \left[\frac{k}{a} + c\lambda_A(H\dot {\phi}^2+\dot \phi\ddot{\phi}) \right]h_A = 0,
\end{align}
with $c=8(2c_1-c_5)$. 
It is clear that only two PV terms, $M_1$ and $M_5$, have nonzero contributions to the evolution of the tensor perturbations, which exhibit the phenomenon of velocity birefringence. More interestingly, when the physical wave number $k/a$ drops below $|c(H\dot {\phi}^2+\dot\phi\ddot{\phi})|$, the frequency squared, $\omega_A^2(t,k)\equiv(k/a)[(k/a)+c\lambda_A(H\dot {\phi}^2+\dot \phi\ddot{\phi})]$, for one of the two polarization states can become negative, potentially leading to a tachyonic instability. When the $\phi$ field serves as the inflaton, and the condition $|c(H\dot {\phi}^2+\dot\phi\ddot{\phi})| > \mathcal{O}(H)$ is satisfied, this transition occurs well inside the horizon during inflation. 
As a result, the mode functions of the corresponding polarization state experience a significant tachyonic instability, leading to an exponential amplification of their amplitude. 
In the next section, we will investigate the primordial GWs generated by the model \eqref{full_action}, assuming $\phi$ acts as the inflaton.

\section{chiral primordial gravitational waves}
\label{sec3}

\begin{figure}[ht]
	\centering
	\includegraphics[width=1\linewidth]{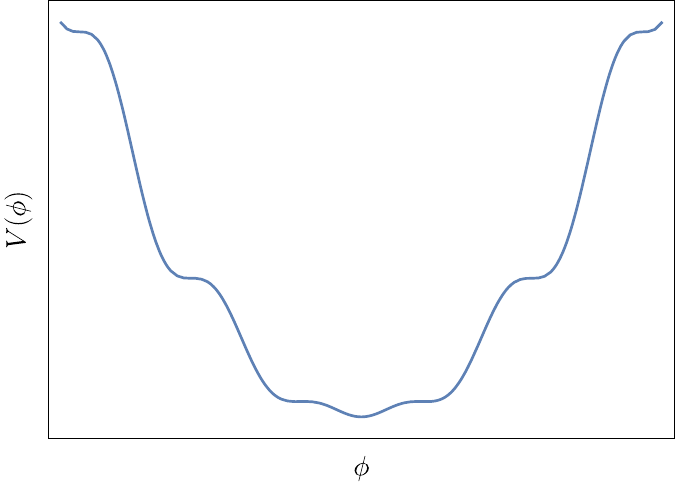}
	\caption{\label{fig1}
		Schematic diagram of the axion potential \eqref{potential}. }
\end{figure}

In this paper, we consider a string-inspired model of axion inflation, in which the inflationary potential takes the following form \cite{Kobayashi:2015aaa,CaboBizet:2016uzv,Ozsoy:2020kat},
\begin{align}\label{potential}
	V(\phi) = \frac{1}{2}m^2\phi^2 + \Lambda^4 \frac{\phi}{f}\sin\left(\frac{\phi}{f}\right).
\end{align}
A salient feature of such a potential is the presence of a step-like structure, arising from the subleading yet significant sinusoidal modulation with $\Lambda^4 \lesssim m^2f^2$. This modulation superimposes steep cliffs and gentle plateaus onto the quadratic potential, as illustrated in Fig. \ref{fig1}. During the period when the inflaton traverses a steep cliff from one plateau to the next, it undergoes rapid acceleration followed by rapid deceleration. As a result, the inflaton velocity exhibits a pronounced peak in its profile. Next, we investigate the impact of the PV term,
as presented in Eq. \eqref{EoM_h_A}, on the tensor perturbations within the nontrivial inflationary dynamics through numerical calculations.

\begin{figure}[ht]
	\centering
	\includegraphics[width=1\linewidth]{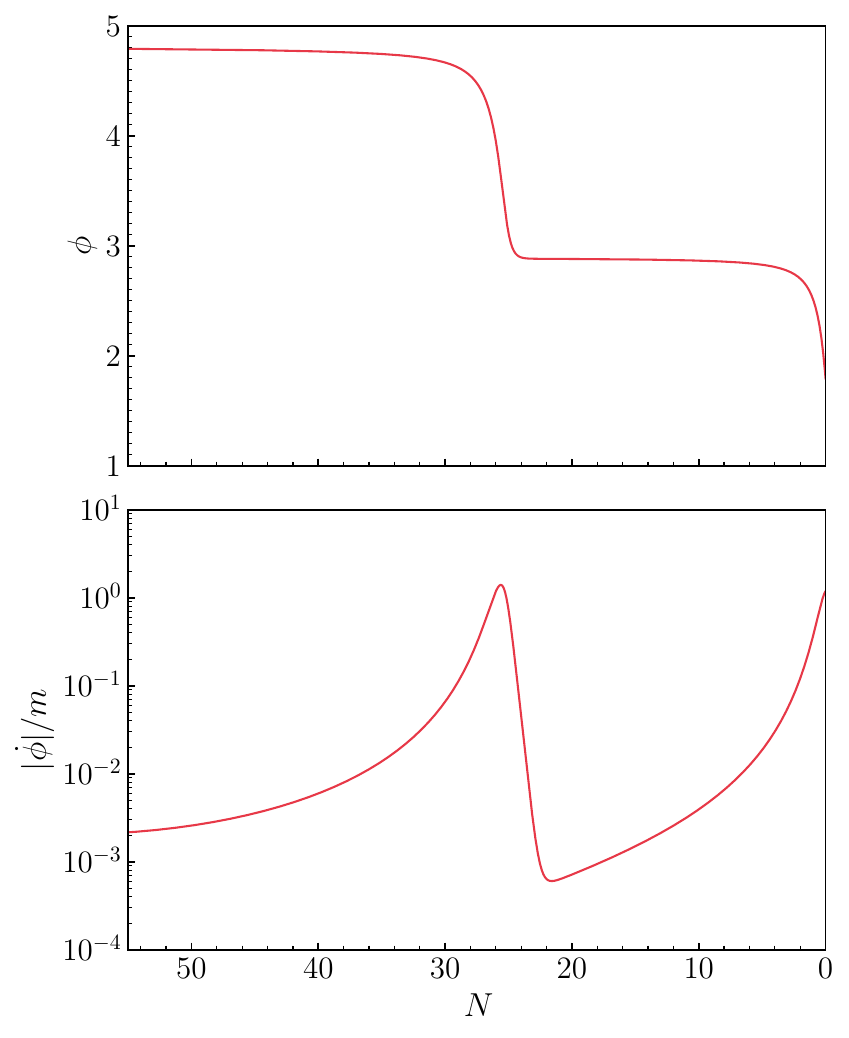}
	\caption{\label{fig2} The evolution of $\phi$ (upper panel) and $|\dot \phi|/m$ (bottom panel) as a function of \textit{e}-folding number $N$.}
\end{figure}

\begin{figure}[ht]
\centering
\includegraphics[width=1\linewidth]{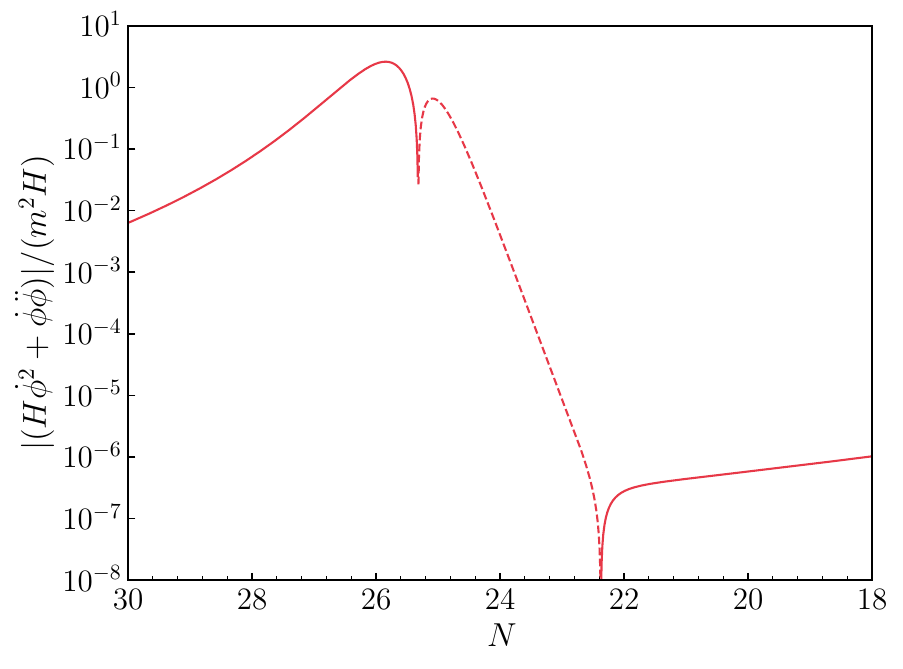}
\caption{\label{fig3}
The evolution of $|{H\dot{\phi}^2 + \dot{\phi}\ddot{\phi}|/(m^2 H)}$ as a function of \textit{e}-folding number $N$. The dashed line denotes the negative value.}
\end{figure}

We adopt the following parameters characterizing the axion potential \eqref{potential}:
\begin{align}\label{para}
	\beta = \frac{\Lambda^4}{m^2f^2}=0.9955,\qquad \alpha = \frac{1}{f}=3.287.
\end{align}
The \textit{e}-folding number is defined as $N \equiv \ln(a_{\rm end}/a)$, where $a_{\rm end}$ denotes the scale factor at the end of inflation. 
We set the \textit{e}-folding number from the time when the pivot scale $k_\ast=0.05\rm{Mpc}^{-1}$ exits the horizon to the end of inflation as $N_\ast=55$. 
By numerically solving the background equations given in Eq. \eqref{BG_Eq}, we present the evolution of the inflaton field and its velocity as functions of the \textit{e}-folding number in Fig. \ref{fig2}. 
It can be seen that the slow-roll evolution of the $\phi$ field across the two smooth plateaus, connected by the steep cliff, occupies a substantial portion of the total 55 \textit{e}-folds. 
The inflaton velocity exhibits a pronounced peak around $N=25$, corresponding to the rapid descent of the inflaton over the steep cliff. 
Since the PV terms do not affect scalar perturbations, the resulting scalar spectral index remains the same as in standard axion inflation. With the parameter choices given in \eqref{para}, the predicted value at pivot scale, $n_s \simeq 0.9622$, is consistent with current CMB observations \cite{BICEP:2021xfz,Planck:2018jri}. 
The parameter $m$ is fixed at $m \simeq 1.6\times10^{-7} M_{\rm Pl}$ by the amplitude of the scalar power spectrum. 
On the other hand, as shown in Fig. \ref{fig3}, which illustrates the evolution of $|{H\dot{\phi}^2 + \dot{\phi}\ddot{\phi}|/(m^2 H)}$, we can take an appropriate coupling constant, such as $c=12~m^{-2}$ in this paper, to ensure that the PV terms have a negligible effect on the tensor perturbation modes at CMB scales. The resulting tensor-to-scalar ratio at pivot scale, $r \simeq 9.5 \times 10^{-6}$, which is nearly identical to that in standard axion inflation, also satisfies the current observational constraints \cite{BICEP:2021xfz,Planck:2018jri}.

\begin{figure}[ht]
	\centering
	\includegraphics[width=1\linewidth]{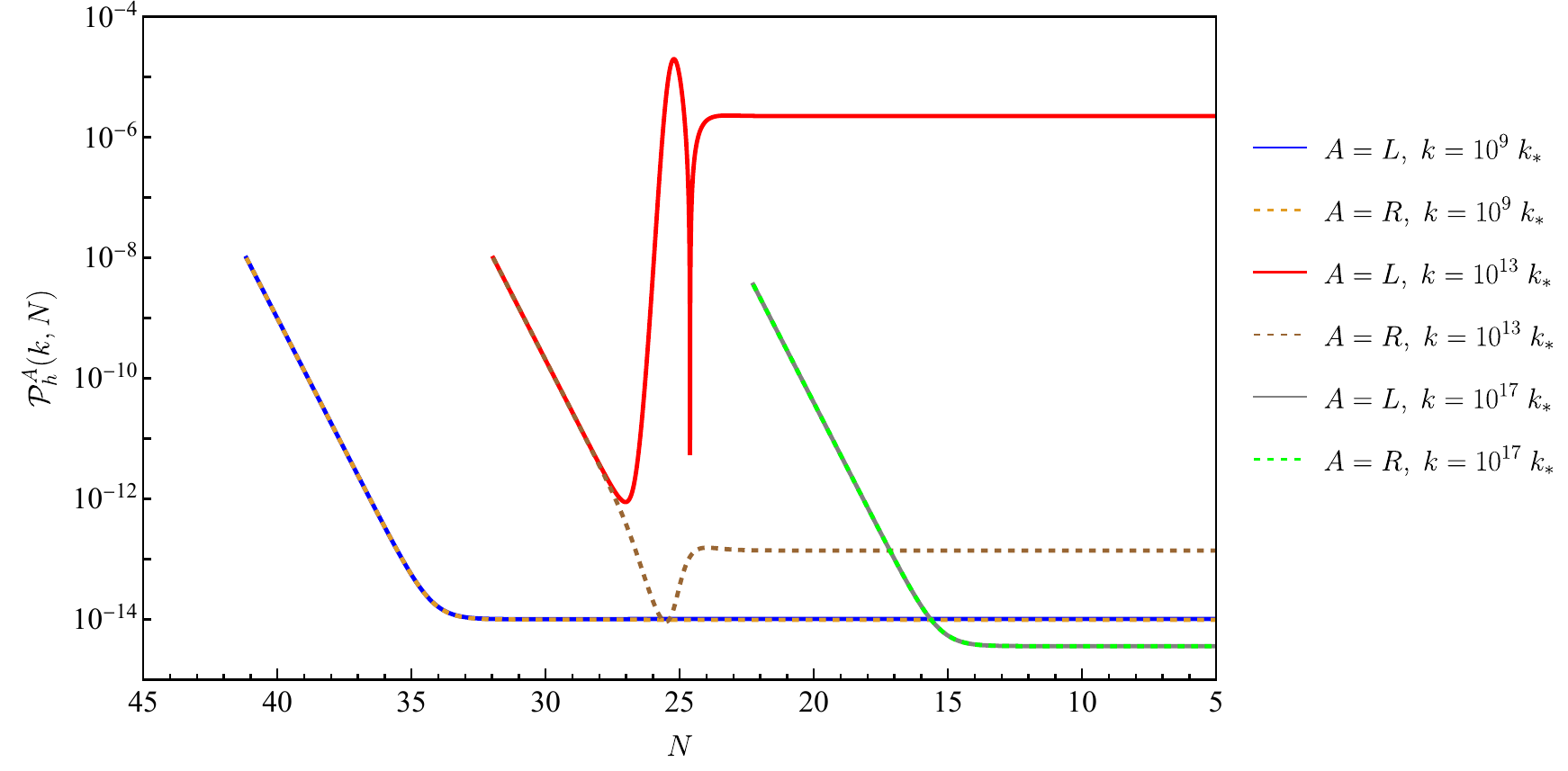}
	\caption{\label{fig4}
		The time evolution of the power spectra $\mathcal{P}^A_h(k,N)$ for different $k$ modes with a coupling constant $c=12~m^{-2}$.}
\end{figure}

\begin{figure}[ht]
	\centering
	\includegraphics[width=1\linewidth]{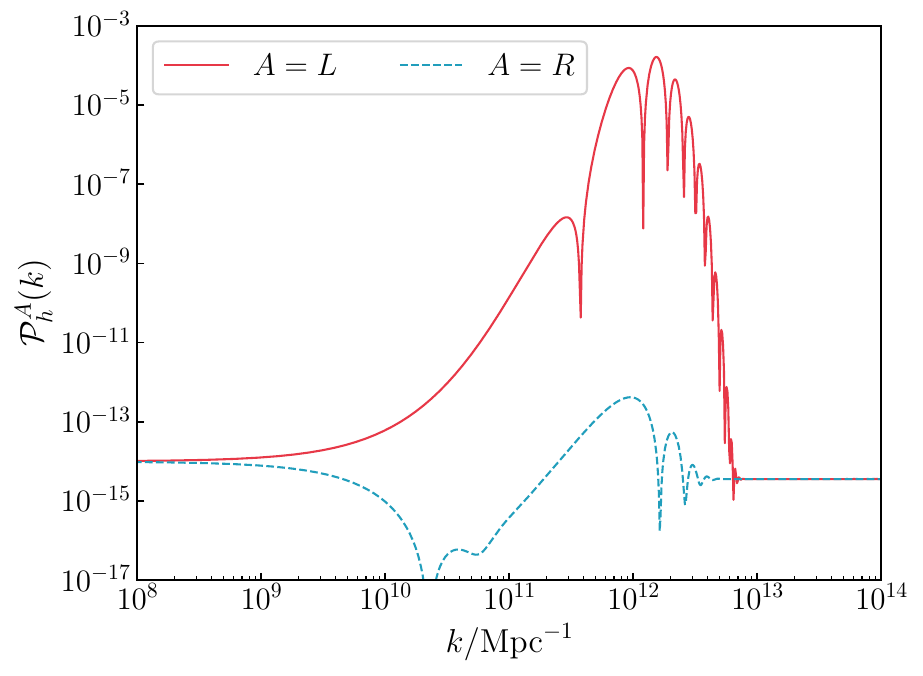}
	\caption{\label{fig5}
		The resulting power spectra for the left- and right-handed polarization states with a coupling constant $c=12~m^{-2}$.}
\end{figure}

It should be noted that the term $({H\dot{\phi}^2 + \dot{\phi}\ddot{\phi})/(m^2 H)}$ exhibits a double-peak structure, one positive and one negative, centered around $N=25$, with the positive peak being more pronounced.
This feature arises from the rapid descent of the inflaton over the steep cliff. In the case of $c=12~m^{-2}$, we observe that $|c(H\dot {\phi}^2+\dot\phi\ddot{\phi})| > \mathcal{O}(H)$ near the peak, suggesting that certain modes experience tachyonic instability. 
In Fig. \ref{fig4}, we show the time evolution of the power spectra $\mathcal{P}^A_h(k,N) \equiv k^3/(2\pi^2) |h_A(k,N)|^2$ for $k = 10^{9}k_{\ast}$, $10^{13}k_{\ast}$, and $10^{17}k_{\ast}$. 
One can observe that the power spectrum with $k=10^{13}k_\ast$ for both the left- and right-handed polarization states undergo growth due to tachyonic instability, but the growth is significantly stronger for the left-handed state and much weaker for the right-handed state. This occurs because, for certain modes (e.g. $k=10^{13}k_\ast$), the condition $\omega_A^2(k)<0$ is satisfied near the peak of the term $({H\dot{\phi}^2 + \dot{\phi}\ddot{\phi})/(m^2 H)}$, i.e., when the $k$ mode is inside the horizon. As a result, the corresponding mode functions experience instability growth. The larger positive peak leads to more significant amplification of the left-handed state, while the smaller negative peak causes relatively weaker amplification of the right-handed state. From Fig. \ref{fig4}, it is also evident that for $k = 10^{9}k_{\ast}$ and $10^{17}k_{\ast}$, the power spectra for both the left- and right-handed polarization states do not experience growth and follow the same evolutionary behavior. This indicates that the evolution of the corresponding mode functions $h_A(k)$ is unaffected by the presence of the PV term. Specifically, these modes are far outside the horizon when $\omega_A^2(k)<0$, and consequently, the mode functions $h_A(k)$ remain constant in this regime, as the frequencies $\omega_A(k)$ are negligible compared to $H$. In conclusion, the power spectra for both polarization states exhibit enhanced amplitude in the range $10^{10}-10^{13}{\rm Mpc}^{-1}$, as shown in Fig. \ref{fig5}. However, the amplitude of the spectrum for the left-handed polarization is significantly larger than that for the right-handed one. It is worth noting that the dominant power spectrum for the left-handed polarization exhibits a distinctive multi-peak structure.

\begin{figure}[ht]
	\centering
	\includegraphics[width=1\linewidth]{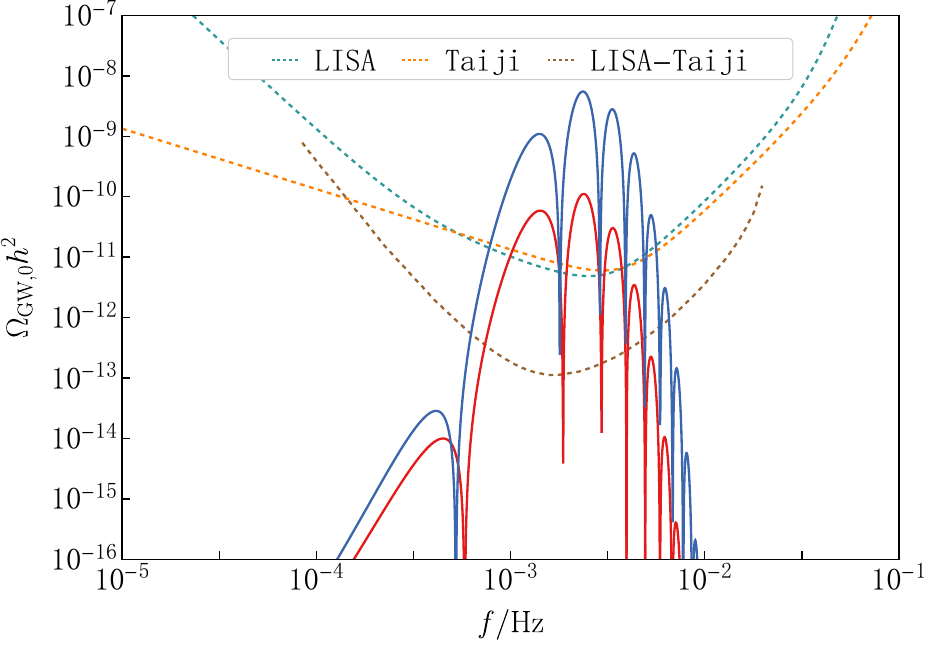}
	\caption{\label{fig6}
		The predicted current energy spectrum of primordial GWs. 	
		The red and blue solid lines represent the energy spectrum for $c = 12~m^{-2}$ and $c = 14~m^{-2}$, respectively. 
		The dashed lines represent the sensitivity curves of LISA (Cyan) \cite{LISA:2017pwj} and Taiji (orange) \cite{Ruan:2018tsw}. The brown dashed line denotes the power-law integrated sensitivity curve of LISA-Taiji network for the $V$ component \cite{Chen:2024ikn}.
	}
\end{figure}

The current GW energy spectrum is related to the primordial tensor power spectrum as follows,
\begin{align}
	\Omega_{\rm GW,0}h^2 &= \left(\Omega^R_{\rm GW,0}+\Omega^L_{\rm GW,0}\right)h^2 \nonumber\\
     &= 6.8\times 10^{-7} \left(\mathcal{P}^R_h(k) + \mathcal{P}^L_h(k)\right).
\end{align}
The observed frequency $f$ is related to the comoving wave number $k$ through $f=1.546\times10^{-15}(k/{\rm Mpc}^{-1}){\rm Hz}$. 
In Fig. \ref{fig6}, we present the current energy spectrum of primordial GWs, $\Omega_{\rm GW,0}h^2$, predicted by the axion inflation within the framework of the PV extensions of STG. 
For $c = 12~m^{-2}$, one can observe that the resulting GW energy spectrum falls within the frequency range accessible to LISA and Taiji, exceeding their sensitivity curves. 
From Fig. \ref{fig5}, it is clear that the observable portion of the GW energy spectrum is dominated by the left-handed polarization. Consequently, this GW signal is almost entirely left-hand polarized, implying that the energy spectrum $\Omega^V_{\rm GW,0}$, related to the Stokes parameter $V$, is nearly identical to the overall energy spectrum $\Omega_{\rm GW,0}$, that is, $\Omega^V_{\rm GW,0} = \Omega^L_{\rm GW,0}-\Omega^R_{\rm GW,0} \simeq \Omega_{\rm GW,0}$.
Recently, Chen \textit{et al}. \cite{Chen:2024ikn} have highlighted the potential for detecting the chirality of a polarized GW background using alternative configurations of the LISA-Taiji network. In Fig. \ref{fig6}, we also present the power-law integrated sensitivity curve of the LISA-Taiji network for the $V$ component of the GW background. The results clearly demonstrate that the chirality of the predicted GW signal could be discerned using the LISA-Taiji network. The fully one-handed polarization of the GW signal discussed in this paper is consistent with that predicted by the inflationary model within the framework of NYmTG. However, the energy spectrum exhibits a distinct multi-peak structure, in contrast to the bump characteristic observed in the NYmTG model. The unique one-handed polarization and distinct energy spectrum profile of primordial GW signal in the model presented in this paper make it distinguishable from other inflationary scenarios.

\section{Fisher Matrix Analysis}
\label{sec4}

In this section, we employ Fisher matrix analysis to estimate the measurement accuracy of the model parameters $\beta$, $\alpha$ and $c$ with the LISA–Taiji network \cite{Chen:2024ikn}. A key prerequisite for achieving optimal parameter estimation is that the peak of the generated GW energy spectrum lies within the most sensitive frequency band of future space-based detectors.
Therefore, it is necessary to investigate how the potential parameters $\alpha$ and $\beta$, which determine the inflationary observables, influence the peak location of the GW spectrum.
%Building on the benchmark parameters, we then utilize the sensitivity curves of LISA, Taiji, and the LISA–Taiji network to determine the lower bound on the coupling parameter $c$ required for producing an observable GW signal.

\begin{figure}[ht]
	\centering
	\includegraphics[width=1\linewidth]{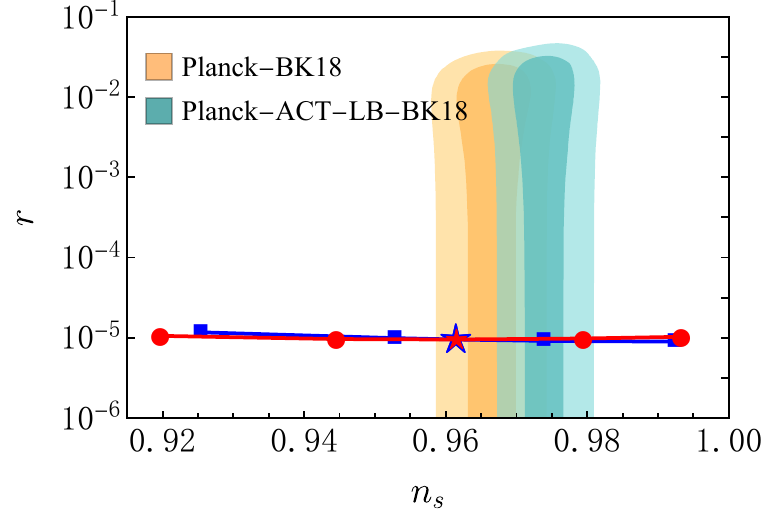}
	\caption{\label{fig7}
		The theoretical predictions in the $n_s - r$ plane for the axion inflation model within the framework of the PV-extended STG. The yellow-shaded region shows the joint constraints from Planck and BICEP/Keck (Planck–BK18); the green-shaded region represents the joint analysis results of Planck and ACT, including CMB lensing effects, BAO measurements from DESI, and the constraints from BICEP/Keck (Planck–ACT–LB–BK18). 
		Specifically, the blue markers correspond to fixed $\beta = 0.9955$, with $\alpha$ taking values from the set $\{3.280, 3.285, 3.287, 3.290, 3.295\}$ from left to right; 
		the red markers correspond to fixed $\alpha = 3.287$, with $\beta$ taking values from the set $\{0.9957, 0.9956, 0.9955, 0.9954, 0.9953\}$ from left to right. 
		The points marked by a star represent the benchmark parameter values (i.e., $\beta=0.9955, \alpha=3.287$) adopted in this paper.
	}
\end{figure}

We begin by examining the impact of the potential parameters $\alpha$ and $\beta$ on inflationary observables, finding that the scalar spectral index $n_s$ exhibits high sensitivity to these variations.
As illustrated in Fig.~\ref{fig7}, we compare the model's predicted trajectories in the $(n_s, r)$ plane under various parameter configurations against the latest observational constraints from Planck–BK18 and Planck–ACT–LB–BK18 \cite{ACT:2025tim}.
Specifically, the blue markers correspond to fixed $\beta = 0.9955$, with $\alpha$ taking values from the set $\{3.280, 3.285, 3.287, 3.290, 3.295\}$ from left to right; whereas the red markers correspond to fixed $\alpha = 3.287$, with $\beta$ taking values from the set $\{0.9957, 0.9956, 0.9955, 0.9954, 0.9953\}$ from left to right. 
The results reveal that when $\beta$ is fixed, $n_s$ increases rapidly as $\alpha$ increases. 
Similarly, when $\alpha$ is fixed, a decrease in $\beta$ leads to a sharp rise in $n_s$. 
Notably, even minor adjustments to $\beta$ result in significant shifts in $n_s$.
In contrast, the impact of these parameters on the tensor-to-scalar ratio $r$ is negligible.
Therefore, by fine-tuning the parameters $\beta$ and $\alpha$, we can effectively adjust the predicted values of $n_s$ and $r$ at the CMB scale, ensuring they fall within the parameter region favored by current observations.

\begin{figure}[ht]
	\centering
	\includegraphics[width=1\linewidth]{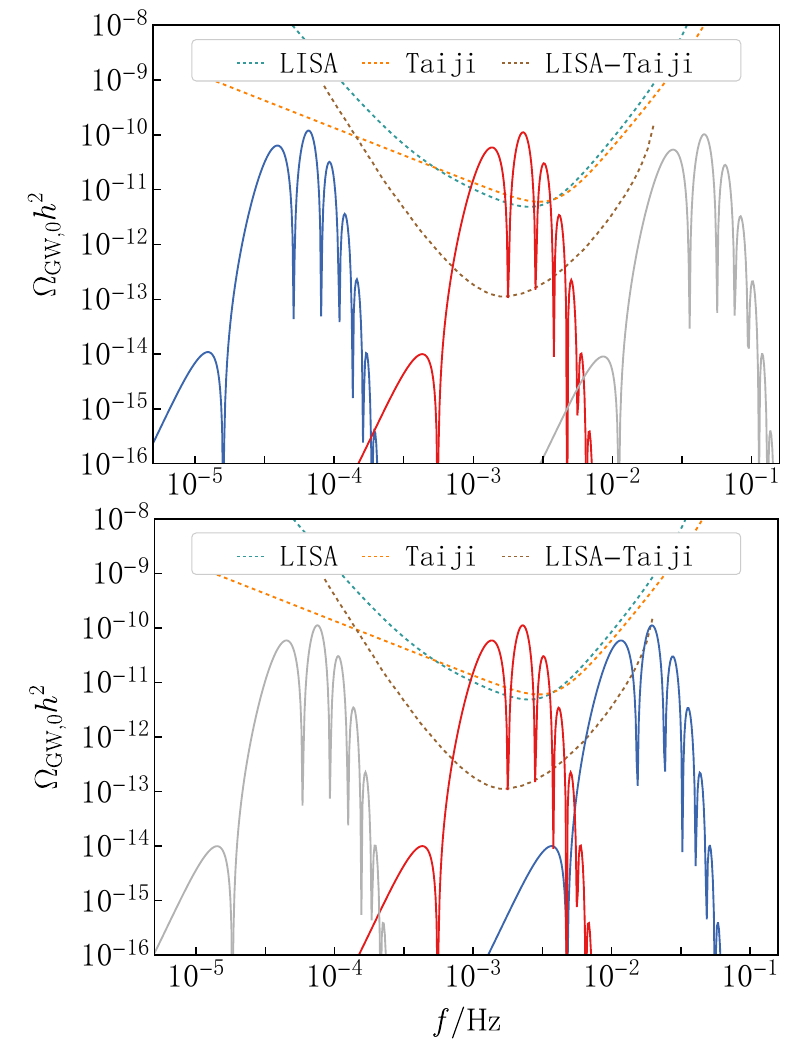}
	\caption{\label{fig8}
		Upper Panel: current GW energy spectra $\Omega_{\rm GW,0}h^2$ for three values of $\alpha$ under fixed parameters $c=12~ m^{-2}$ and $\beta=0.9955$. The curves, from left to right, correspond to $\alpha = 3.280$ (blue solid line), $\alpha = 3.287$ (red solid line), and $\alpha = 3.295$ (gray solid line).
		Bottom Panel: current GW energy spectra $\Omega_{\rm GW,0}h^2$ for three values of $\beta$ under fixed parameters $c=12~m^{-2}$ and $\alpha=3.287$. The curves, from left to right, correspond to $\beta = 0.9957$ (gray solid line), $\beta = 0.9955$ (red solid line), and $\beta = 0.9953$ (blue solid line).
		The dashed lines represent the sensitivity curves of LISA (Cyan), Taiji (orange), and LISA-Taiji network (brown). 
	}
\end{figure}

In addition to satisfying the CMB constraints, our study reveals that the peak position of the primordial GW energy spectrum is significantly constrained by the potential parameters $\alpha$ and $\beta$.
As shown in Fig. \ref{fig8}, we compute the current GW energy spectra by varying $\alpha$ and $\beta$ individually while keeping the coupling $c$ fixed. 
Specifically, the top panel corresponds to the case with fixed $\beta$ and $\alpha=\{3.280, 3.287, 3.295\}$,
observing that the GW energy spectrum undergoes a pronounced horizontal shift toward higher frequencies as $\alpha$ increases. 
Similarly, the bottom panel displays the spectra for $\beta=\{0.9957, 0.9955, 0.9953\}$ with fixed $\alpha$, demonstrating that a decrease in $\beta$ also drives the spectrum toward higher frequencies.
Therefore, combining these considerations, we adopted $\beta=0.9955$ and $\alpha=3.287$ as the benchmark values in our previous analysis. 
This specific parameter set not only satisfies the observational constraints on $n_s$ and $r$ but also ensures that the peak position of the generated GW spectrum falls precisely within the most sensitive region of future GW detectors.

To assess the precision of parameter estimation, we perform the widely-used Fisher matrix analysis \cite{Vallisneri:2007ev} using the fiducial values of the potential parameters established above. The Fisher matrix expression for the GW model parameters is given by \cite{Su:2025nkl}
\begin{align}
    F_{ab}&=4T_{\rm obs}\left( \frac{3H^2_0}{4\pi^2}\right)^2\times \nonumber \\
    &\sum_{\kappa}\int^\infty_0 {\rm d}f\frac{(\Gamma^I_\kappa+\Pi\,\Gamma^V_\kappa)^2\partial_{\theta_a}\Omega_{\rm GW}(f)\partial_{\theta_b}\Omega_{\rm GW}(f)}{f^6 N^2_\kappa},
\end{align}
where $\theta_a$ and $\theta_b$ refer to arbitrary model parameters. Here, $N_\kappa$ is the signal variance caused by noise, $\Gamma^I_\kappa$ and $\Gamma^V_\kappa$ are the overlap reduction functions for the intensity and circular-polarization components, which can be found in \cite{Chen:2024ikn}, and $\kappa$ labels the data channel pairs \cite{Chen:2024ikn}. The Hubble constant is $H_0=h\,100\,{\rm km\,s^{-1}\,Mpc^{-1}}$ with $h=0.67$. The effective observation time is denoted by $T_{\rm obs}$, which is set to 3 years in this work. The circular polarization parameter $\Pi\equiv(\mathcal{P}_h^L-\mathcal{P}_h^R)/(\mathcal{P}_h^L+\mathcal{P}_h^R)$ is almost close to unity in our model. Although the GW energy spectrum for $c=12~m^{-2}$ is potentially detectable (see Fig.~\ref{fig6}), a larger signal amplitude is required to place meaningful constraints on the model parameters. As an illustration, for $c=14~m^{-2}$, we obtain at the $1\sigma$ confidence level, 
\begin{align}
&\beta=0.9955\pm0.0109~(1.1\%), \;\;\alpha=3.287 \pm 0.3292~(10\%), \nonumber\\
&c\,m^2 = 14\pm3.5637~(25.4\%),
\end{align}
where the percentages in parentheses denote the relative errors. Thus, we find that the parameter $\beta$ can be determined with relatively high precision, whereas the measurement precision of $c$ is significantly lower compared to $\beta$. This is likely because the GW spectrum is more sensitive to variations in $\beta$ than to those in $c$.

\section{conclusions}
\label{conclusion}

The PV extensions of STG introduce the PV interactions between the scalar field and parity-odd terms quadratic in the nonmetricity tensor. These PV terms do not have effects on the evolution of the background and linear scalar perturbations. However, the PV terms produce the velocity birefringence in the tensor perturbations, potentially leading to intriguing phenomenological consequences if such a phenomenon occurs during inflation. In this paper, we apply the PV extensions of STG to axion inflation and investigate their effect on the primordial tensor perturbations. During inflation, the rapid descent of the inflaton over the steep cliff of its potential generates a pronounced peak in the inflaton velocity profile. Consequently, the PV term that appears in the equation of motion for tensor perturbations exhibits a double-peak structure, one positive and one negative, with the positive peak being more pronounced. The large positive peak of the PV term induces tachyonic instability in tensor modes at specific scales for the left-handed polarization state, resulting in an exponential amplification of their amplitudes. 
At approximately the same scales, tensor modes for the right-handed polarization state also experience growth due to the tachyonic instability caused by the smaller negative peak. However, this growth is negligible compared to that of the left-handed polarization state. 
As a result, the predicted chiral GW signal, dominated by the left-handed polarization state, exhibits a significant amplitude potentially detectable by LISA and Taiji.
Moreover, the chirality of the signal could be identified using the LISA-Taiji network. 
The unique one-handed polarization, together with the characteristic multi-peak structure of the GW energy spectrum, sets this signal apart from other stochastic GW backgrounds.

Furthermore, we use the Fisher matrix to estimate the measurement precision of the potential parameters $\beta$ and $\alpha$, as well as the coupling constant $c$, with the LISA-Taiji network. We choose the potential parameters $\beta$ and $\alpha$ to satisfy current CMB observational constraints, while ensuring that the peak of the generated GW spectrum falls within the optimal sensitivity window of space-based detectors. With an appropriate choice of $c$ that yields a sufficiently large GW energy spectrum, we find that $\beta$ can be measured with relatively high precision, whereas the constraint on $c$ is substantially weaker than this on $\beta$.

In summary, the axion inflation scenario presented here provides a clear phenomenological pathway to test parity violation on cosmological scales. 
The potential detection of GW signals with distinct chiral features would serve as a powerful tool for probing inflation dynamics and testing STG theories.

%%%%%%%%%%%%%%%%%%%%%%%%%%%%%%%%%%%%%%%%%%%%%%%%%%%%%%%%%%%%%%%%%%%%%%%%%%%%%%%%
%%%%%%%%%%%%%%%%%%%%%%%%%%%%%%  %%%%%%%%%%%%%%%%%%%%%%%%%%%%%%%%%
%%%%%%%%%%%%%%%%%%%%%%%%%%%%%%%%%%%%%%%%%%%%%%%%%%%%%%%%%%%%%%%%%%%%%%%%%%%%%%%%
\begin{acknowledgments}
We appreciate very much the insightful comments and helpful suggestions by the anonymous referee. We thank Chang Liu for stimulating discussions. This work is supported in part by the National Natural Science Foundation of China Grants No. 12305057, No. 12375045, No.~12275080 and No.~12075084, and the Innovative Research Group of Hunan Province (Grant No.~2024JJ1006).

\end{acknowledgments}

%%%%%%%%%%%%%%%%%%%%%%%%%%%%%%%%%%%%%%%%%%%%%%%%%%%%%%%%%%%%%%%%%%%%%%%%%%%%%%%%
%%%%%%%%%%%%%%%%%%%%%%%%%%%%%%%%%%%%% references %%%%%%%%%%%%%%%%%%%%%%%%%%%%%%%%%%%%
%%%%%%%%%%%%%%%%%%%%%%%%%%%%%%%%%%%%%%%%%%%%%%%%%%%%%%%%%%%%%%%%%%%%%%%%%%%%%%%%
\bibliographystyle{apsrev4-1}
\bibliography{ref}

%merlin.mbs apsrev4-1.bst 2010-07-25 4.21a (PWD, AO, DPC) hacked
%Control: key (0)
%Control: author (72) initials jnrlst
%Control: editor formatted (1) identically to author
%Control: production of article title (-1) disabled
%Control: page (0) single
%Control: year (1) truncated
%Control: production of eprint (0) enabled
\begin{thebibliography}{62}%
\makeatletter
\providecommand \@ifxundefined [1]{%
 \@ifx{#1\undefined}
}%
\providecommand \@ifnum [1]{%
 \ifnum #1\expandafter \@firstoftwo
 \else \expandafter \@secondoftwo
 \fi
}%
\providecommand \@ifx [1]{%
 \ifx #1\expandafter \@firstoftwo
 \else \expandafter \@secondoftwo
 \fi
}%
\providecommand \natexlab [1]{#1}%
\providecommand \enquote  [1]{``#1''}%
\providecommand \bibnamefont  [1]{#1}%
\providecommand \bibfnamefont [1]{#1}%
\providecommand \citenamefont [1]{#1}%
\providecommand \href@noop [0]{\@secondoftwo}%
\providecommand \href [0]{\begingroup \@sanitize@url \@href}%
\providecommand \@href[1]{\@@startlink{#1}\@@href}%
\providecommand \@@href[1]{\endgroup#1\@@endlink}%
\providecommand \@sanitize@url [0]{\catcode `\\12\catcode `\$12\catcode
  `\&12\catcode `\#12\catcode `\^12\catcode `\_12\catcode `\%12\relax}%
\providecommand \@@startlink[1]{}%
\providecommand \@@endlink[0]{}%
\providecommand \url  [0]{\begingroup\@sanitize@url \@url }%
\providecommand \@url [1]{\endgroup\@href {#1}{\urlprefix }}%
\providecommand \urlprefix  [0]{URL }%
\providecommand \Eprint [0]{\href }%
\providecommand \doibase [0]{http://dx.doi.org/}%
\providecommand \selectlanguage [0]{\@gobble}%
\providecommand \bibinfo  [0]{\@secondoftwo}%
\providecommand \bibfield  [0]{\@secondoftwo}%
\providecommand \translation [1]{[#1]}%
\providecommand \BibitemOpen [0]{}%
\providecommand \bibitemStop [0]{}%
\providecommand \bibitemNoStop [0]{.\EOS\space}%
\providecommand \EOS [0]{\spacefactor3000\relax}%
\providecommand \BibitemShut  [1]{\csname bibitem#1\endcsname}%
\let\auto@bib@innerbib\@empty
%</preamble>
\bibitem [{\citenamefont {Abbott}\ \emph
  {et~al.}(2016{\natexlab{a}})\citenamefont {Abbott} \emph
  {et~al.}}]{LIGOScientific:2016sjg}%
  \BibitemOpen
  \bibfield  {author} {\bibinfo {author} {\bibfnamefont {B.~P.}\ \bibnamefont
  {Abbott}} \emph {et~al.} (\bibinfo {collaboration} {LIGO Scientific,
  Virgo}),\ }\href {\doibase 10.1103/PhysRevLett.116.241103} {\bibfield
  {journal} {\bibinfo  {journal} {Phys. Rev. Lett.}\ }\textbf {\bibinfo
  {volume} {116}},\ \bibinfo {pages} {241103} (\bibinfo {year}
  {2016}{\natexlab{a}})}\BibitemShut {NoStop}%
\bibitem [{\citenamefont {Abbott}\ \emph
  {et~al.}(2016{\natexlab{b}})\citenamefont {Abbott} \emph
  {et~al.}}]{LIGOScientific:2016aoc}%
  \BibitemOpen
  \bibfield  {author} {\bibinfo {author} {\bibfnamefont {B.~P.}\ \bibnamefont
  {Abbott}} \emph {et~al.} (\bibinfo {collaboration} {LIGO Scientific,
  Virgo}),\ }\href {\doibase 10.1103/PhysRevLett.116.061102} {\bibfield
  {journal} {\bibinfo  {journal} {Phys. Rev. Lett.}\ }\textbf {\bibinfo
  {volume} {116}},\ \bibinfo {pages} {061102} (\bibinfo {year}
  {2016}{\natexlab{b}})}\BibitemShut {NoStop}%
\bibitem [{\citenamefont {Abbott}\ \emph
  {et~al.}(2017{\natexlab{a}})\citenamefont {Abbott} \emph
  {et~al.}}]{LIGOScientific:2017vox}%
  \BibitemOpen
  \bibfield  {author} {\bibinfo {author} {\bibfnamefont {B.~P.}\ \bibnamefont
  {Abbott}} \emph {et~al.} (\bibinfo {collaboration} {LIGO Scientific,
  Virgo}),\ }\href {\doibase 10.3847/2041-8213/aa9f0c} {\bibfield  {journal}
  {\bibinfo  {journal} {Astrophys. J. Lett.}\ }\textbf {\bibinfo {volume}
  {851}},\ \bibinfo {pages} {L35} (\bibinfo {year}
  {2017}{\natexlab{a}})}\BibitemShut {NoStop}%
\bibitem [{\citenamefont {Abbott}\ \emph
  {et~al.}(2017{\natexlab{b}})\citenamefont {Abbott} \emph
  {et~al.}}]{LIGOScientific:2017vwq}%
  \BibitemOpen
  \bibfield  {author} {\bibinfo {author} {\bibfnamefont {B.~P.}\ \bibnamefont
  {Abbott}} \emph {et~al.} (\bibinfo {collaboration} {LIGO Scientific,
  Virgo}),\ }\href {\doibase 10.1103/PhysRevLett.119.161101} {\bibfield
  {journal} {\bibinfo  {journal} {Phys. Rev. Lett.}\ }\textbf {\bibinfo
  {volume} {119}},\ \bibinfo {pages} {161101} (\bibinfo {year}
  {2017}{\natexlab{b}})}\BibitemShut {NoStop}%
\bibitem [{\citenamefont {Abbott}\ \emph
  {et~al.}(2017{\natexlab{c}})\citenamefont {Abbott} \emph
  {et~al.}}]{LIGOScientific:2017ycc}%
  \BibitemOpen
  \bibfield  {author} {\bibinfo {author} {\bibfnamefont {B.~P.}\ \bibnamefont
  {Abbott}} \emph {et~al.} (\bibinfo {collaboration} {LIGO Scientific,
  Virgo}),\ }\href {\doibase 10.1103/PhysRevLett.119.141101} {\bibfield
  {journal} {\bibinfo  {journal} {Phys. Rev. Lett.}\ }\textbf {\bibinfo
  {volume} {119}},\ \bibinfo {pages} {141101} (\bibinfo {year}
  {2017}{\natexlab{c}})}\BibitemShut {NoStop}%
\bibitem [{\citenamefont {Abbott}\ \emph {et~al.}(2019)\citenamefont {Abbott}
  \emph {et~al.}}]{LIGOScientific:2018mvr}%
  \BibitemOpen
  \bibfield  {author} {\bibinfo {author} {\bibfnamefont {B.~P.}\ \bibnamefont
  {Abbott}} \emph {et~al.} (\bibinfo {collaboration} {LIGO Scientific,
  Virgo}),\ }\href {\doibase 10.1103/PhysRevX.9.031040} {\bibfield  {journal}
  {\bibinfo  {journal} {Phys. Rev. X}\ }\textbf {\bibinfo {volume} {9}},\
  \bibinfo {pages} {031040} (\bibinfo {year} {2019})}\BibitemShut {NoStop}%
\bibitem [{\citenamefont {Abbott}\ \emph {et~al.}(2020)\citenamefont {Abbott}
  \emph {et~al.}}]{LIGOScientific:2020aai}%
  \BibitemOpen
  \bibfield  {author} {\bibinfo {author} {\bibfnamefont {B.~P.}\ \bibnamefont
  {Abbott}} \emph {et~al.} (\bibinfo {collaboration} {LIGO Scientific,
  Virgo}),\ }\href {\doibase 10.3847/2041-8213/ab75f5} {\bibfield  {journal}
  {\bibinfo  {journal} {Astrophys. J. Lett.}\ }\textbf {\bibinfo {volume}
  {892}},\ \bibinfo {pages} {L3} (\bibinfo {year} {2020})}\BibitemShut
  {NoStop}%
\bibitem [{\citenamefont {Guzzetti}\ \emph {et~al.}(2016)\citenamefont
  {Guzzetti}, \citenamefont {Bartolo}, \citenamefont {Liguori},\ and\
  \citenamefont {Matarrese}}]{Guzzetti:2016mkm}%
  \BibitemOpen
  \bibfield  {author} {\bibinfo {author} {\bibfnamefont {M.~C.}\ \bibnamefont
  {Guzzetti}}, \bibinfo {author} {\bibfnamefont {N.}~\bibnamefont {Bartolo}},
  \bibinfo {author} {\bibfnamefont {M.}~\bibnamefont {Liguori}}, \ and\
  \bibinfo {author} {\bibfnamefont {S.}~\bibnamefont {Matarrese}},\ }\href
  {\doibase 10.1393/ncr/i2016-10127-1} {\bibfield  {journal} {\bibinfo
  {journal} {Riv. Nuovo Cim.}\ }\textbf {\bibinfo {volume} {39}},\ \bibinfo
  {pages} {399} (\bibinfo {year} {2016})},\ \Eprint
  {http://arxiv.org/abs/1605.01615} {arXiv:1605.01615 [astro-ph.CO]}
  \BibitemShut {NoStop}%
\bibitem [{\citenamefont {Bartolo}\ \emph {et~al.}(2016)\citenamefont {Bartolo}
  \emph {et~al.}}]{Bartolo:2016ami}%
  \BibitemOpen
  \bibfield  {author} {\bibinfo {author} {\bibfnamefont {N.}~\bibnamefont
  {Bartolo}} \emph {et~al.},\ }\href {\doibase 10.1088/1475-7516/2016/12/026}
  {\bibfield  {journal} {\bibinfo  {journal} {JCAP}\ }\textbf {\bibinfo
  {volume} {12}},\ \bibinfo {pages} {026} (\bibinfo {year} {2016})},\ \Eprint
  {http://arxiv.org/abs/1610.06481} {arXiv:1610.06481 [astro-ph.CO]}
  \BibitemShut {NoStop}%
\bibitem [{\citenamefont {Auclair}\ \emph {et~al.}(2023)\citenamefont {Auclair}
  \emph {et~al.}}]{LISACosmologyWorkingGroup:2022jok}%
  \BibitemOpen
  \bibfield  {author} {\bibinfo {author} {\bibfnamefont {P.}~\bibnamefont
  {Auclair}} \emph {et~al.} (\bibinfo {collaboration} {LISA Cosmology Working
  Group}),\ }\href {\doibase 10.1007/s41114-023-00045-2} {\bibfield  {journal}
  {\bibinfo  {journal} {Living Rev. Rel.}\ }\textbf {\bibinfo {volume} {26}},\
  \bibinfo {pages} {5} (\bibinfo {year} {2023})},\ \Eprint
  {http://arxiv.org/abs/2204.05434} {arXiv:2204.05434 [astro-ph.CO]}
  \BibitemShut {NoStop}%
\bibitem [{\citenamefont {Cai}\ \emph {et~al.}(2017)\citenamefont {Cai},
  \citenamefont {Cao}, \citenamefont {Guo}, \citenamefont {Wang},\ and\
  \citenamefont {Yang}}]{Cai:2017cbj}%
  \BibitemOpen
  \bibfield  {author} {\bibinfo {author} {\bibfnamefont {R.-G.}\ \bibnamefont
  {Cai}}, \bibinfo {author} {\bibfnamefont {Z.}~\bibnamefont {Cao}}, \bibinfo
  {author} {\bibfnamefont {Z.-K.}\ \bibnamefont {Guo}}, \bibinfo {author}
  {\bibfnamefont {S.-J.}\ \bibnamefont {Wang}}, \ and\ \bibinfo {author}
  {\bibfnamefont {T.}~\bibnamefont {Yang}},\ }\href {\doibase
  10.1093/nsr/nwx029} {\bibfield  {journal} {\bibinfo  {journal} {Natl. Sci.
  Rev.}\ }\textbf {\bibinfo {volume} {4}},\ \bibinfo {pages} {687} (\bibinfo
  {year} {2017})},\ \Eprint {http://arxiv.org/abs/1703.00187} {arXiv:1703.00187
  [gr-qc]} \BibitemShut {NoStop}%
\bibitem [{\citenamefont {Caprini}\ and\ \citenamefont
  {Figueroa}(2018)}]{Caprini:2018mtu}%
  \BibitemOpen
  \bibfield  {author} {\bibinfo {author} {\bibfnamefont {C.}~\bibnamefont
  {Caprini}}\ and\ \bibinfo {author} {\bibfnamefont {D.~G.}\ \bibnamefont
  {Figueroa}},\ }\href {\doibase 10.1088/1361-6382/aac608} {\bibfield
  {journal} {\bibinfo  {journal} {Class. Quant. Grav.}\ }\textbf {\bibinfo
  {volume} {35}},\ \bibinfo {pages} {163001} (\bibinfo {year} {2018})},\
  \Eprint {http://arxiv.org/abs/1801.04268} {arXiv:1801.04268 [astro-ph.CO]}
  \BibitemShut {NoStop}%
\bibitem [{\citenamefont {Kuroyanagi}\ \emph {et~al.}(2018)\citenamefont
  {Kuroyanagi}, \citenamefont {Chiba},\ and\ \citenamefont
  {Takahashi}}]{Kuroyanagi:2018csn}%
  \BibitemOpen
  \bibfield  {author} {\bibinfo {author} {\bibfnamefont {S.}~\bibnamefont
  {Kuroyanagi}}, \bibinfo {author} {\bibfnamefont {T.}~\bibnamefont {Chiba}}, \
  and\ \bibinfo {author} {\bibfnamefont {T.}~\bibnamefont {Takahashi}},\ }\href
  {\doibase 10.1088/1475-7516/2018/11/038} {\bibfield  {journal} {\bibinfo
  {journal} {JCAP}\ }\textbf {\bibinfo {volume} {11}},\ \bibinfo {pages} {038}
  (\bibinfo {year} {2018})},\ \Eprint {http://arxiv.org/abs/1807.00786}
  {arXiv:1807.00786 [astro-ph.CO]} \BibitemShut {NoStop}%
\bibitem [{\citenamefont {Ade}\ \emph {et~al.}(2021)\citenamefont {Ade} \emph
  {et~al.}}]{BICEP:2021xfz}%
  \BibitemOpen
  \bibfield  {author} {\bibinfo {author} {\bibfnamefont {P.~A.~R.}\
  \bibnamefont {Ade}} \emph {et~al.} (\bibinfo {collaboration} {BICEP, Keck}),\
  }\href {\doibase 10.1103/PhysRevLett.127.151301} {\bibfield  {journal}
  {\bibinfo  {journal} {Phys. Rev. Lett.}\ }\textbf {\bibinfo {volume} {127}},\
  \bibinfo {pages} {151301} (\bibinfo {year} {2021})}\BibitemShut {NoStop}%
\bibitem [{\citenamefont {Amaro-Seoane}\ \emph {et~al.}(2017)\citenamefont
  {Amaro-Seoane} \emph {et~al.}}]{LISA:2017pwj}%
  \BibitemOpen
  \bibfield  {author} {\bibinfo {author} {\bibfnamefont {P.}~\bibnamefont
  {Amaro-Seoane}} \emph {et~al.} (\bibinfo {collaboration} {LISA}),\
  }\href@noop {} {\  (\bibinfo {year} {2017})},\ \Eprint
  {http://arxiv.org/abs/1702.00786} {arXiv:1702.00786 [astro-ph.IM]}
  \BibitemShut {NoStop}%
\bibitem [{\citenamefont {Barausse}\ \emph {et~al.}(2020)\citenamefont
  {Barausse} \emph {et~al.}}]{Barausse:2020rsu}%
  \BibitemOpen
  \bibfield  {author} {\bibinfo {author} {\bibfnamefont {E.}~\bibnamefont
  {Barausse}} \emph {et~al.},\ }\href {\doibase 10.1007/s10714-020-02691-1}
  {\bibfield  {journal} {\bibinfo  {journal} {Gen. Rel. Grav.}\ }\textbf
  {\bibinfo {volume} {52}},\ \bibinfo {pages} {81} (\bibinfo {year} {2020})},\
  \Eprint {http://arxiv.org/abs/2001.09793} {arXiv:2001.09793 [gr-qc]}
  \BibitemShut {NoStop}%
\bibitem [{\citenamefont {Hu}\ and\ \citenamefont {Wu}(2017)}]{Hu:2017mde}%
  \BibitemOpen
  \bibfield  {author} {\bibinfo {author} {\bibfnamefont {W.-R.}\ \bibnamefont
  {Hu}}\ and\ \bibinfo {author} {\bibfnamefont {Y.-L.}\ \bibnamefont {Wu}},\
  }\href {\doibase 10.1093/nsr/nwx116} {\bibfield  {journal} {\bibinfo
  {journal} {Natl. Sci. Rev.}\ }\textbf {\bibinfo {volume} {4}},\ \bibinfo
  {pages} {685} (\bibinfo {year} {2017})}\BibitemShut {NoStop}%
\bibitem [{\citenamefont {Ruan}\ \emph {et~al.}(2020)\citenamefont {Ruan},
  \citenamefont {Guo}, \citenamefont {Cai},\ and\ \citenamefont
  {Zhang}}]{Ruan:2018tsw}%
  \BibitemOpen
  \bibfield  {author} {\bibinfo {author} {\bibfnamefont {W.-H.}\ \bibnamefont
  {Ruan}}, \bibinfo {author} {\bibfnamefont {Z.-K.}\ \bibnamefont {Guo}},
  \bibinfo {author} {\bibfnamefont {R.-G.}\ \bibnamefont {Cai}}, \ and\
  \bibinfo {author} {\bibfnamefont {Y.-Z.}\ \bibnamefont {Zhang}},\ }\href
  {\doibase 10.1142/S0217751X2050075X} {\bibfield  {journal} {\bibinfo
  {journal} {Int. J. Mod. Phys. A}\ }\textbf {\bibinfo {volume} {35}},\
  \bibinfo {pages} {2050075} (\bibinfo {year} {2020})},\ \Eprint
  {http://arxiv.org/abs/1807.09495} {arXiv:1807.09495 [gr-qc]} \BibitemShut
  {NoStop}%
\bibitem [{\citenamefont {Cannone}\ \emph {et~al.}(2015)\citenamefont
  {Cannone}, \citenamefont {Tasinato},\ and\ \citenamefont
  {Wands}}]{Cannone:2014uqa}%
  \BibitemOpen
  \bibfield  {author} {\bibinfo {author} {\bibfnamefont {D.}~\bibnamefont
  {Cannone}}, \bibinfo {author} {\bibfnamefont {G.}~\bibnamefont {Tasinato}}, \
  and\ \bibinfo {author} {\bibfnamefont {D.}~\bibnamefont {Wands}},\ }\href
  {\doibase 10.1088/1475-7516/2015/01/029} {\bibfield  {journal} {\bibinfo
  {journal} {JCAP}\ }\textbf {\bibinfo {volume} {01}},\ \bibinfo {pages} {029}
  (\bibinfo {year} {2015})},\ \Eprint {http://arxiv.org/abs/1409.6568}
  {arXiv:1409.6568 [astro-ph.CO]} \BibitemShut {NoStop}%
\bibitem [{\citenamefont {Mylova}\ \emph {et~al.}(2018)\citenamefont {Mylova},
  \citenamefont {{\"O}zsoy}, \citenamefont {Parameswaran}, \citenamefont
  {Tasinato},\ and\ \citenamefont {Zavala}}]{Mylova:2018yap}%
  \BibitemOpen
  \bibfield  {author} {\bibinfo {author} {\bibfnamefont {M.}~\bibnamefont
  {Mylova}}, \bibinfo {author} {\bibfnamefont {O.}~\bibnamefont {{\"O}zsoy}},
  \bibinfo {author} {\bibfnamefont {S.}~\bibnamefont {Parameswaran}}, \bibinfo
  {author} {\bibfnamefont {G.}~\bibnamefont {Tasinato}}, \ and\ \bibinfo
  {author} {\bibfnamefont {I.}~\bibnamefont {Zavala}},\ }\href {\doibase
  10.1088/1475-7516/2018/12/024} {\bibfield  {journal} {\bibinfo  {journal}
  {JCAP}\ }\textbf {\bibinfo {volume} {12}},\ \bibinfo {pages} {024} (\bibinfo
  {year} {2018})},\ \Eprint {http://arxiv.org/abs/1808.10475} {arXiv:1808.10475
  [gr-qc]} \BibitemShut {NoStop}%
\bibitem [{\citenamefont {Ozsoy}\ \emph {et~al.}(2019)\citenamefont {Ozsoy},
  \citenamefont {Mylova}, \citenamefont {Parameswaran}, \citenamefont {Powell},
  \citenamefont {Tasinato},\ and\ \citenamefont {Zavala}}]{Ozsoy:2019slf}%
  \BibitemOpen
  \bibfield  {author} {\bibinfo {author} {\bibfnamefont {O.}~\bibnamefont
  {Ozsoy}}, \bibinfo {author} {\bibfnamefont {M.}~\bibnamefont {Mylova}},
  \bibinfo {author} {\bibfnamefont {S.}~\bibnamefont {Parameswaran}}, \bibinfo
  {author} {\bibfnamefont {C.}~\bibnamefont {Powell}}, \bibinfo {author}
  {\bibfnamefont {G.}~\bibnamefont {Tasinato}}, \ and\ \bibinfo {author}
  {\bibfnamefont {I.}~\bibnamefont {Zavala}},\ }\href {\doibase
  10.1088/1475-7516/2019/09/036} {\bibfield  {journal} {\bibinfo  {journal}
  {JCAP}\ }\textbf {\bibinfo {volume} {09}},\ \bibinfo {pages} {036} (\bibinfo
  {year} {2019})},\ \Eprint {http://arxiv.org/abs/1902.04976} {arXiv:1902.04976
  [hep-th]} \BibitemShut {NoStop}%
\bibitem [{\citenamefont {Lue}\ \emph {et~al.}(1999)\citenamefont {Lue},
  \citenamefont {Wang},\ and\ \citenamefont {Kamionkowski}}]{Lue:1998mq}%
  \BibitemOpen
  \bibfield  {author} {\bibinfo {author} {\bibfnamefont {A.}~\bibnamefont
  {Lue}}, \bibinfo {author} {\bibfnamefont {L.-M.}\ \bibnamefont {Wang}}, \
  and\ \bibinfo {author} {\bibfnamefont {M.}~\bibnamefont {Kamionkowski}},\
  }\href {\doibase 10.1103/PhysRevLett.83.1506} {\bibfield  {journal} {\bibinfo
   {journal} {Phys. Rev. Lett.}\ }\textbf {\bibinfo {volume} {83}},\ \bibinfo
  {pages} {1506} (\bibinfo {year} {1999})},\ \Eprint
  {http://arxiv.org/abs/astro-ph/9812088} {arXiv:astro-ph/9812088} \BibitemShut
  {NoStop}%
\bibitem [{\citenamefont {Jackiw}\ and\ \citenamefont
  {Pi}(2003)}]{Jackiw:2003pm}%
  \BibitemOpen
  \bibfield  {author} {\bibinfo {author} {\bibfnamefont {R.}~\bibnamefont
  {Jackiw}}\ and\ \bibinfo {author} {\bibfnamefont {S.~Y.}\ \bibnamefont
  {Pi}},\ }\href {\doibase 10.1103/PhysRevD.68.104012} {\bibfield  {journal}
  {\bibinfo  {journal} {Phys. Rev. D}\ }\textbf {\bibinfo {volume} {68}},\
  \bibinfo {pages} {104012} (\bibinfo {year} {2003})}\BibitemShut {NoStop}%
\bibitem [{\citenamefont {Alexander}\ and\ \citenamefont
  {Yunes}(2009)}]{Alexander:2009tp}%
  \BibitemOpen
  \bibfield  {author} {\bibinfo {author} {\bibfnamefont {S.}~\bibnamefont
  {Alexander}}\ and\ \bibinfo {author} {\bibfnamefont {N.}~\bibnamefont
  {Yunes}},\ }\href {\doibase 10.1016/j.physrep.2009.07.002} {\bibfield
  {journal} {\bibinfo  {journal} {Phys. Rept.}\ }\textbf {\bibinfo {volume}
  {480}},\ \bibinfo {pages} {1} (\bibinfo {year} {2009})}\BibitemShut {NoStop}%
\bibitem [{\citenamefont {Crisostomi}\ \emph {et~al.}(2018)\citenamefont
  {Crisostomi}, \citenamefont {Noui}, \citenamefont {Charmousis},\ and\
  \citenamefont {Langlois}}]{Crisostomi:2017ugk}%
  \BibitemOpen
  \bibfield  {author} {\bibinfo {author} {\bibfnamefont {M.}~\bibnamefont
  {Crisostomi}}, \bibinfo {author} {\bibfnamefont {K.}~\bibnamefont {Noui}},
  \bibinfo {author} {\bibfnamefont {C.}~\bibnamefont {Charmousis}}, \ and\
  \bibinfo {author} {\bibfnamefont {D.}~\bibnamefont {Langlois}},\ }\href
  {\doibase 10.1103/PhysRevD.97.044034} {\bibfield  {journal} {\bibinfo
  {journal} {Phys. Rev. D}\ }\textbf {\bibinfo {volume} {97}},\ \bibinfo
  {pages} {044034} (\bibinfo {year} {2018})},\ \Eprint
  {http://arxiv.org/abs/1710.04531} {arXiv:1710.04531 [hep-th]} \BibitemShut
  {NoStop}%
\bibitem [{\citenamefont {Nishizawa}\ and\ \citenamefont
  {Kobayashi}(2018)}]{Nishizawa:2018srh}%
  \BibitemOpen
  \bibfield  {author} {\bibinfo {author} {\bibfnamefont {A.}~\bibnamefont
  {Nishizawa}}\ and\ \bibinfo {author} {\bibfnamefont {T.}~\bibnamefont
  {Kobayashi}},\ }\href {\doibase 10.1103/PhysRevD.98.124018} {\bibfield
  {journal} {\bibinfo  {journal} {Phys. Rev. D}\ }\textbf {\bibinfo {volume}
  {98}},\ \bibinfo {pages} {124018} (\bibinfo {year} {2018})},\ \Eprint
  {http://arxiv.org/abs/1809.00815} {arXiv:1809.00815 [gr-qc]} \BibitemShut
  {NoStop}%
\bibitem [{\citenamefont {Gao}\ and\ \citenamefont {Hong}(2020)}]{Gao:2019liu}%
  \BibitemOpen
  \bibfield  {author} {\bibinfo {author} {\bibfnamefont {X.}~\bibnamefont
  {Gao}}\ and\ \bibinfo {author} {\bibfnamefont {X.-Y.}\ \bibnamefont {Hong}},\
  }\href {\doibase 10.1103/PhysRevD.101.064057} {\bibfield  {journal} {\bibinfo
   {journal} {Phys. Rev. D}\ }\textbf {\bibinfo {volume} {101}},\ \bibinfo
  {pages} {064057} (\bibinfo {year} {2020})},\ \Eprint
  {http://arxiv.org/abs/1906.07131} {arXiv:1906.07131 [gr-qc]} \BibitemShut
  {NoStop}%
\bibitem [{\citenamefont {Li}\ \emph {et~al.}(2020)\citenamefont {Li},
  \citenamefont {Rao},\ and\ \citenamefont {Zhao}}]{Li:2020xjt}%
  \BibitemOpen
  \bibfield  {author} {\bibinfo {author} {\bibfnamefont {M.}~\bibnamefont
  {Li}}, \bibinfo {author} {\bibfnamefont {H.}~\bibnamefont {Rao}}, \ and\
  \bibinfo {author} {\bibfnamefont {D.}~\bibnamefont {Zhao}},\ }\href {\doibase
  10.1088/1475-7516/2020/11/023} {\bibfield  {journal} {\bibinfo  {journal}
  {JCAP}\ }\textbf {\bibinfo {volume} {11}},\ \bibinfo {pages} {023} (\bibinfo
  {year} {2020})},\ \Eprint {http://arxiv.org/abs/2007.08038} {arXiv:2007.08038
  [gr-qc]} \BibitemShut {NoStop}%
\bibitem [{\citenamefont {Li}\ \emph {et~al.}(2022{\natexlab{a}})\citenamefont
  {Li}, \citenamefont {Li},\ and\ \citenamefont {Rao}}]{Li:2022mti}%
  \BibitemOpen
  \bibfield  {author} {\bibinfo {author} {\bibfnamefont {M.}~\bibnamefont
  {Li}}, \bibinfo {author} {\bibfnamefont {Z.}~\bibnamefont {Li}}, \ and\
  \bibinfo {author} {\bibfnamefont {H.}~\bibnamefont {Rao}},\ }\href {\doibase
  10.1016/j.physletb.2022.137395} {\bibfield  {journal} {\bibinfo  {journal}
  {Phys. Lett. B}\ }\textbf {\bibinfo {volume} {834}},\ \bibinfo {pages}
  {137395} (\bibinfo {year} {2022}{\natexlab{a}})},\ \Eprint
  {http://arxiv.org/abs/2201.02357} {arXiv:2201.02357 [gr-qc]} \BibitemShut
  {NoStop}%
\bibitem [{\citenamefont {Conroy}\ and\ \citenamefont
  {Koivisto}(2019)}]{Conroy:2019ibo}%
  \BibitemOpen
  \bibfield  {author} {\bibinfo {author} {\bibfnamefont {A.}~\bibnamefont
  {Conroy}}\ and\ \bibinfo {author} {\bibfnamefont {T.}~\bibnamefont
  {Koivisto}},\ }\href {\doibase 10.1088/1475-7516/2019/12/016} {\bibfield
  {journal} {\bibinfo  {journal} {JCAP}\ }\textbf {\bibinfo {volume} {12}},\
  \bibinfo {pages} {016} (\bibinfo {year} {2019})},\ \Eprint
  {http://arxiv.org/abs/1908.04313} {arXiv:1908.04313 [gr-qc]} \BibitemShut
  {NoStop}%
\bibitem [{\citenamefont {Li}\ and\ \citenamefont {Zhao}(2022)}]{Li:2021mdp}%
  \BibitemOpen
  \bibfield  {author} {\bibinfo {author} {\bibfnamefont {M.}~\bibnamefont
  {Li}}\ and\ \bibinfo {author} {\bibfnamefont {D.}~\bibnamefont {Zhao}},\
  }\href {\doibase 10.1016/j.physletb.2022.136968} {\bibfield  {journal}
  {\bibinfo  {journal} {Phys. Lett. B}\ }\textbf {\bibinfo {volume} {827}},\
  \bibinfo {pages} {136968} (\bibinfo {year} {2022})},\ \Eprint
  {http://arxiv.org/abs/2108.01337} {arXiv:2108.01337 [gr-qc]} \BibitemShut
  {NoStop}%
\bibitem [{\citenamefont {Li}\ \emph {et~al.}(2022{\natexlab{b}})\citenamefont
  {Li}, \citenamefont {Tong},\ and\ \citenamefont {Zhao}}]{Li:2022vtn}%
  \BibitemOpen
  \bibfield  {author} {\bibinfo {author} {\bibfnamefont {M.}~\bibnamefont
  {Li}}, \bibinfo {author} {\bibfnamefont {Y.}~\bibnamefont {Tong}}, \ and\
  \bibinfo {author} {\bibfnamefont {D.}~\bibnamefont {Zhao}},\ }\href {\doibase
  10.1103/PhysRevD.105.104002} {\bibfield  {journal} {\bibinfo  {journal}
  {Phys. Rev. D}\ }\textbf {\bibinfo {volume} {105}},\ \bibinfo {pages}
  {104002} (\bibinfo {year} {2022}{\natexlab{b}})},\ \Eprint
  {http://arxiv.org/abs/2203.06912} {arXiv:2203.06912 [gr-qc]} \BibitemShut
  {NoStop}%
\bibitem [{\citenamefont {Alexander}\ \emph {et~al.}(2006)\citenamefont
  {Alexander}, \citenamefont {Peskin},\ and\ \citenamefont
  {Sheikh-Jabbari}}]{Alexander:2004us}%
  \BibitemOpen
  \bibfield  {author} {\bibinfo {author} {\bibfnamefont {S.~H.-S.}\
  \bibnamefont {Alexander}}, \bibinfo {author} {\bibfnamefont {M.~E.}\
  \bibnamefont {Peskin}}, \ and\ \bibinfo {author} {\bibfnamefont {M.~M.}\
  \bibnamefont {Sheikh-Jabbari}},\ }\href {\doibase
  10.1103/PhysRevLett.96.081301} {\bibfield  {journal} {\bibinfo  {journal}
  {Phys. Rev. Lett.}\ }\textbf {\bibinfo {volume} {96}},\ \bibinfo {pages}
  {081301} (\bibinfo {year} {2006})},\ \Eprint
  {http://arxiv.org/abs/hep-th/0403069} {arXiv:hep-th/0403069} \BibitemShut
  {NoStop}%
\bibitem [{\citenamefont {Satoh}\ \emph {et~al.}(2008)\citenamefont {Satoh},
  \citenamefont {Kanno},\ and\ \citenamefont {Soda}}]{Satoh:2007gn}%
  \BibitemOpen
  \bibfield  {author} {\bibinfo {author} {\bibfnamefont {M.}~\bibnamefont
  {Satoh}}, \bibinfo {author} {\bibfnamefont {S.}~\bibnamefont {Kanno}}, \ and\
  \bibinfo {author} {\bibfnamefont {J.}~\bibnamefont {Soda}},\ }\href {\doibase
  10.1103/PhysRevD.77.023526} {\bibfield  {journal} {\bibinfo  {journal} {Phys.
  Rev. D}\ }\textbf {\bibinfo {volume} {77}},\ \bibinfo {pages} {023526}
  (\bibinfo {year} {2008})},\ \Eprint {http://arxiv.org/abs/0706.3585}
  {arXiv:0706.3585 [astro-ph]} \BibitemShut {NoStop}%
\bibitem [{\citenamefont {Saito}\ \emph {et~al.}(2007)\citenamefont {Saito},
  \citenamefont {Ichiki},\ and\ \citenamefont {Taruya}}]{Saito:2007kt}%
  \BibitemOpen
  \bibfield  {author} {\bibinfo {author} {\bibfnamefont {S.}~\bibnamefont
  {Saito}}, \bibinfo {author} {\bibfnamefont {K.}~\bibnamefont {Ichiki}}, \
  and\ \bibinfo {author} {\bibfnamefont {A.}~\bibnamefont {Taruya}},\ }\href
  {\doibase 10.1088/1475-7516/2007/09/002} {\bibfield  {journal} {\bibinfo
  {journal} {JCAP}\ }\textbf {\bibinfo {volume} {09}},\ \bibinfo {pages} {002}
  (\bibinfo {year} {2007})},\ \Eprint {http://arxiv.org/abs/0705.3701}
  {arXiv:0705.3701 [astro-ph]} \BibitemShut {NoStop}%
\bibitem [{\citenamefont {Gluscevic}\ and\ \citenamefont
  {Kamionkowski}(2010)}]{Gluscevic:2010vv}%
  \BibitemOpen
  \bibfield  {author} {\bibinfo {author} {\bibfnamefont {V.}~\bibnamefont
  {Gluscevic}}\ and\ \bibinfo {author} {\bibfnamefont {M.}~\bibnamefont
  {Kamionkowski}},\ }\href {\doibase 10.1103/PhysRevD.81.123529} {\bibfield
  {journal} {\bibinfo  {journal} {Phys. Rev. D}\ }\textbf {\bibinfo {volume}
  {81}},\ \bibinfo {pages} {123529} (\bibinfo {year} {2010})},\ \Eprint
  {http://arxiv.org/abs/1002.1308} {arXiv:1002.1308 [astro-ph.CO]} \BibitemShut
  {NoStop}%
\bibitem [{\citenamefont {Wang}\ \emph {et~al.}(2013)\citenamefont {Wang},
  \citenamefont {Wu}, \citenamefont {Zhao},\ and\ \citenamefont
  {Zhu}}]{Wang:2012fi}%
  \BibitemOpen
  \bibfield  {author} {\bibinfo {author} {\bibfnamefont {A.}~\bibnamefont
  {Wang}}, \bibinfo {author} {\bibfnamefont {Q.}~\bibnamefont {Wu}}, \bibinfo
  {author} {\bibfnamefont {W.}~\bibnamefont {Zhao}}, \ and\ \bibinfo {author}
  {\bibfnamefont {T.}~\bibnamefont {Zhu}},\ }\href {\doibase
  10.1103/PhysRevD.87.103512} {\bibfield  {journal} {\bibinfo  {journal} {Phys.
  Rev. D}\ }\textbf {\bibinfo {volume} {87}},\ \bibinfo {pages} {103512}
  (\bibinfo {year} {2013})},\ \Eprint {http://arxiv.org/abs/1208.5490}
  {arXiv:1208.5490 [astro-ph.CO]} \BibitemShut {NoStop}%
\bibitem [{\citenamefont {Alexander}(2016)}]{Alexander:2016hxk}%
  \BibitemOpen
  \bibfield  {author} {\bibinfo {author} {\bibfnamefont {S.~H.~S.}\
  \bibnamefont {Alexander}},\ }\href {\doibase 10.1142/S0218271816400137}
  {\bibfield  {journal} {\bibinfo  {journal} {Int. J. Mod. Phys. D}\ }\textbf
  {\bibinfo {volume} {25}},\ \bibinfo {pages} {1640013} (\bibinfo {year}
  {2016})},\ \Eprint {http://arxiv.org/abs/1604.00703} {arXiv:1604.00703
  [hep-th]} \BibitemShut {NoStop}%
\bibitem [{\citenamefont {Bartolo}\ and\ \citenamefont
  {Orlando}(2017)}]{Bartolo:2017szm}%
  \BibitemOpen
  \bibfield  {author} {\bibinfo {author} {\bibfnamefont {N.}~\bibnamefont
  {Bartolo}}\ and\ \bibinfo {author} {\bibfnamefont {G.}~\bibnamefont
  {Orlando}},\ }\href {\doibase 10.1088/1475-7516/2017/07/034} {\bibfield
  {journal} {\bibinfo  {journal} {JCAP}\ }\textbf {\bibinfo {volume} {07}},\
  \bibinfo {pages} {034} (\bibinfo {year} {2017})},\ \Eprint
  {http://arxiv.org/abs/1706.04627} {arXiv:1706.04627 [astro-ph.CO]}
  \BibitemShut {NoStop}%
\bibitem [{\citenamefont {Qiao}\ \emph {et~al.}(2020)\citenamefont {Qiao},
  \citenamefont {Zhu}, \citenamefont {Zhao},\ and\ \citenamefont
  {Wang}}]{Qiao:2019hkz}%
  \BibitemOpen
  \bibfield  {author} {\bibinfo {author} {\bibfnamefont {J.}~\bibnamefont
  {Qiao}}, \bibinfo {author} {\bibfnamefont {T.}~\bibnamefont {Zhu}}, \bibinfo
  {author} {\bibfnamefont {W.}~\bibnamefont {Zhao}}, \ and\ \bibinfo {author}
  {\bibfnamefont {A.}~\bibnamefont {Wang}},\ }\href {\doibase
  10.1103/PhysRevD.101.043528} {\bibfield  {journal} {\bibinfo  {journal}
  {Phys. Rev. D}\ }\textbf {\bibinfo {volume} {101}},\ \bibinfo {pages}
  {043528} (\bibinfo {year} {2020})},\ \Eprint
  {http://arxiv.org/abs/1911.01580} {arXiv:1911.01580 [astro-ph.CO]}
  \BibitemShut {NoStop}%
\bibitem [{\citenamefont {Wang}\ \emph {et~al.}(2021)\citenamefont {Wang},
  \citenamefont {Niu}, \citenamefont {Zhu},\ and\ \citenamefont
  {Zhao}}]{Wang:2020cub}%
  \BibitemOpen
  \bibfield  {author} {\bibinfo {author} {\bibfnamefont {Y.-F.}\ \bibnamefont
  {Wang}}, \bibinfo {author} {\bibfnamefont {R.}~\bibnamefont {Niu}}, \bibinfo
  {author} {\bibfnamefont {T.}~\bibnamefont {Zhu}}, \ and\ \bibinfo {author}
  {\bibfnamefont {W.}~\bibnamefont {Zhao}},\ }\href {\doibase
  10.3847/1538-4357/abd7a6} {\bibfield  {journal} {\bibinfo  {journal}
  {Astrophys. J.}\ }\textbf {\bibinfo {volume} {908}},\ \bibinfo {pages} {58}
  (\bibinfo {year} {2021})},\ \Eprint {http://arxiv.org/abs/2002.05668}
  {arXiv:2002.05668 [gr-qc]} \BibitemShut {NoStop}%
\bibitem [{\citenamefont {Okounkova}\ \emph {et~al.}(2022)\citenamefont
  {Okounkova}, \citenamefont {Farr}, \citenamefont {Isi},\ and\ \citenamefont
  {Stein}}]{Okounkova:2021xjv}%
  \BibitemOpen
  \bibfield  {author} {\bibinfo {author} {\bibfnamefont {M.}~\bibnamefont
  {Okounkova}}, \bibinfo {author} {\bibfnamefont {W.~M.}\ \bibnamefont {Farr}},
  \bibinfo {author} {\bibfnamefont {M.}~\bibnamefont {Isi}}, \ and\ \bibinfo
  {author} {\bibfnamefont {L.~C.}\ \bibnamefont {Stein}},\ }\href {\doibase
  10.1103/PhysRevD.106.044067} {\bibfield  {journal} {\bibinfo  {journal}
  {Phys. Rev. D}\ }\textbf {\bibinfo {volume} {106}},\ \bibinfo {pages}
  {044067} (\bibinfo {year} {2022})},\ \Eprint
  {http://arxiv.org/abs/2101.11153} {arXiv:2101.11153 [gr-qc]} \BibitemShut
  {NoStop}%
\bibitem [{\citenamefont {Wu}\ \emph {et~al.}(2022)\citenamefont {Wu},
  \citenamefont {Zhu}, \citenamefont {Niu}, \citenamefont {Zhao},\ and\
  \citenamefont {Wang}}]{Wu:2021ndf}%
  \BibitemOpen
  \bibfield  {author} {\bibinfo {author} {\bibfnamefont {Q.}~\bibnamefont
  {Wu}}, \bibinfo {author} {\bibfnamefont {T.}~\bibnamefont {Zhu}}, \bibinfo
  {author} {\bibfnamefont {R.}~\bibnamefont {Niu}}, \bibinfo {author}
  {\bibfnamefont {W.}~\bibnamefont {Zhao}}, \ and\ \bibinfo {author}
  {\bibfnamefont {A.}~\bibnamefont {Wang}},\ }\href {\doibase
  10.1103/PhysRevD.105.024035} {\bibfield  {journal} {\bibinfo  {journal}
  {Phys. Rev. D}\ }\textbf {\bibinfo {volume} {105}},\ \bibinfo {pages}
  {024035} (\bibinfo {year} {2022})},\ \Eprint
  {http://arxiv.org/abs/2110.13870} {arXiv:2110.13870 [gr-qc]} \BibitemShut
  {NoStop}%
\bibitem [{\citenamefont {Gong}\ \emph {et~al.}(2022)\citenamefont {Gong},
  \citenamefont {Zhu}, \citenamefont {Niu}, \citenamefont {Wu}, \citenamefont
  {Cui}, \citenamefont {Zhang}, \citenamefont {Zhao},\ and\ \citenamefont
  {Wang}}]{Gong:2021jgg}%
  \BibitemOpen
  \bibfield  {author} {\bibinfo {author} {\bibfnamefont {C.}~\bibnamefont
  {Gong}}, \bibinfo {author} {\bibfnamefont {T.}~\bibnamefont {Zhu}}, \bibinfo
  {author} {\bibfnamefont {R.}~\bibnamefont {Niu}}, \bibinfo {author}
  {\bibfnamefont {Q.}~\bibnamefont {Wu}}, \bibinfo {author} {\bibfnamefont
  {J.-L.}\ \bibnamefont {Cui}}, \bibinfo {author} {\bibfnamefont
  {X.}~\bibnamefont {Zhang}}, \bibinfo {author} {\bibfnamefont
  {W.}~\bibnamefont {Zhao}}, \ and\ \bibinfo {author} {\bibfnamefont
  {A.}~\bibnamefont {Wang}},\ }\href {\doibase 10.1103/PhysRevD.105.044034}
  {\bibfield  {journal} {\bibinfo  {journal} {Phys. Rev. D}\ }\textbf {\bibinfo
  {volume} {105}},\ \bibinfo {pages} {044034} (\bibinfo {year} {2022})},\
  \Eprint {http://arxiv.org/abs/2112.06446} {arXiv:2112.06446 [gr-qc]}
  \BibitemShut {NoStop}%
\bibitem [{\citenamefont {Odintsov}\ and\ \citenamefont
  {Oikonomou}(2022)}]{Odintsov:2022hxu}%
  \BibitemOpen
  \bibfield  {author} {\bibinfo {author} {\bibfnamefont {S.~D.}\ \bibnamefont
  {Odintsov}}\ and\ \bibinfo {author} {\bibfnamefont {V.~K.}\ \bibnamefont
  {Oikonomou}},\ }\href {\doibase 10.1103/PhysRevD.105.104054} {\bibfield
  {journal} {\bibinfo  {journal} {Phys. Rev. D}\ }\textbf {\bibinfo {volume}
  {105}},\ \bibinfo {pages} {104054} (\bibinfo {year} {2022})},\ \Eprint
  {http://arxiv.org/abs/2205.07304} {arXiv:2205.07304 [gr-qc]} \BibitemShut
  {NoStop}%
\bibitem [{\citenamefont {Zhu}\ \emph {et~al.}(2023{\natexlab{a}})\citenamefont
  {Zhu}, \citenamefont {Zhao},\ and\ \citenamefont {Wang}}]{Zhu:2022uoq}%
  \BibitemOpen
  \bibfield  {author} {\bibinfo {author} {\bibfnamefont {T.}~\bibnamefont
  {Zhu}}, \bibinfo {author} {\bibfnamefont {W.}~\bibnamefont {Zhao}}, \ and\
  \bibinfo {author} {\bibfnamefont {A.}~\bibnamefont {Wang}},\ }\href {\doibase
  10.1103/PhysRevD.107.044051} {\bibfield  {journal} {\bibinfo  {journal}
  {Phys. Rev. D}\ }\textbf {\bibinfo {volume} {107}},\ \bibinfo {pages}
  {044051} (\bibinfo {year} {2023}{\natexlab{a}})},\ \Eprint
  {http://arxiv.org/abs/2211.04711} {arXiv:2211.04711 [gr-qc]} \BibitemShut
  {NoStop}%
\bibitem [{\citenamefont {Zhu}\ \emph {et~al.}(2023{\natexlab{b}})\citenamefont
  {Zhu}, \citenamefont {Zhao},\ and\ \citenamefont {Wang}}]{Zhu:2022dfq}%
  \BibitemOpen
  \bibfield  {author} {\bibinfo {author} {\bibfnamefont {T.}~\bibnamefont
  {Zhu}}, \bibinfo {author} {\bibfnamefont {W.}~\bibnamefont {Zhao}}, \ and\
  \bibinfo {author} {\bibfnamefont {A.}~\bibnamefont {Wang}},\ }\href {\doibase
  10.1103/PhysRevD.107.024031} {\bibfield  {journal} {\bibinfo  {journal}
  {Phys. Rev. D}\ }\textbf {\bibinfo {volume} {107}},\ \bibinfo {pages}
  {024031} (\bibinfo {year} {2023}{\natexlab{b}})},\ \Eprint
  {http://arxiv.org/abs/2210.05259} {arXiv:2210.05259 [gr-qc]} \BibitemShut
  {NoStop}%
\bibitem [{\citenamefont {Fu}\ \emph {et~al.}(2021)\citenamefont {Fu},
  \citenamefont {Liu}, \citenamefont {Zhu}, \citenamefont {Yu},\ and\
  \citenamefont {Wu}}]{Fu:2020tlw}%
  \BibitemOpen
  \bibfield  {author} {\bibinfo {author} {\bibfnamefont {C.}~\bibnamefont
  {Fu}}, \bibinfo {author} {\bibfnamefont {J.}~\bibnamefont {Liu}}, \bibinfo
  {author} {\bibfnamefont {T.}~\bibnamefont {Zhu}}, \bibinfo {author}
  {\bibfnamefont {H.}~\bibnamefont {Yu}}, \ and\ \bibinfo {author}
  {\bibfnamefont {P.}~\bibnamefont {Wu}},\ }\href {\doibase
  10.1140/epjc/s10052-021-09001-2} {\bibfield  {journal} {\bibinfo  {journal}
  {Eur. Phys. J. C}\ }\textbf {\bibinfo {volume} {81}},\ \bibinfo {pages} {204}
  (\bibinfo {year} {2021})},\ \Eprint {http://arxiv.org/abs/2006.03771}
  {arXiv:2006.03771 [gr-qc]} \BibitemShut {NoStop}%
\bibitem [{\citenamefont {Peng}\ \emph {et~al.}(2022)\citenamefont {Peng},
  \citenamefont {Zeng}, \citenamefont {Fu},\ and\ \citenamefont
  {Guo}}]{Peng:2022ttg}%
  \BibitemOpen
  \bibfield  {author} {\bibinfo {author} {\bibfnamefont {Z.-Z.}\ \bibnamefont
  {Peng}}, \bibinfo {author} {\bibfnamefont {Z.-M.}\ \bibnamefont {Zeng}},
  \bibinfo {author} {\bibfnamefont {C.}~\bibnamefont {Fu}}, \ and\ \bibinfo
  {author} {\bibfnamefont {Z.-K.}\ \bibnamefont {Guo}},\ }\href {\doibase
  10.1103/PhysRevD.106.124044} {\bibfield  {journal} {\bibinfo  {journal}
  {Phys. Rev. D}\ }\textbf {\bibinfo {volume} {106}},\ \bibinfo {pages}
  {124044} (\bibinfo {year} {2022})},\ \Eprint
  {http://arxiv.org/abs/2209.10374} {arXiv:2209.10374 [gr-qc]} \BibitemShut
  {NoStop}%
\bibitem [{\citenamefont {Cai}\ \emph {et~al.}(2022)\citenamefont {Cai},
  \citenamefont {Fu},\ and\ \citenamefont {Yu}}]{Cai:2021uup}%
  \BibitemOpen
  \bibfield  {author} {\bibinfo {author} {\bibfnamefont {R.-G.}\ \bibnamefont
  {Cai}}, \bibinfo {author} {\bibfnamefont {C.}~\bibnamefont {Fu}}, \ and\
  \bibinfo {author} {\bibfnamefont {W.-W.}\ \bibnamefont {Yu}},\ }\href
  {\doibase 10.1103/PhysRevD.105.103520} {\bibfield  {journal} {\bibinfo
  {journal} {Phys. Rev. D}\ }\textbf {\bibinfo {volume} {105}},\ \bibinfo
  {pages} {103520} (\bibinfo {year} {2022})},\ \Eprint
  {http://arxiv.org/abs/2112.04794} {arXiv:2112.04794 [astro-ph.CO]}
  \BibitemShut {NoStop}%
\bibitem [{\citenamefont {Fu}\ \emph {et~al.}(2024)\citenamefont {Fu},
  \citenamefont {Liu}, \citenamefont {Yang}, \citenamefont {Yu},\ and\
  \citenamefont {Zhang}}]{Fu:2023aab}%
  \BibitemOpen
  \bibfield  {author} {\bibinfo {author} {\bibfnamefont {C.}~\bibnamefont
  {Fu}}, \bibinfo {author} {\bibfnamefont {J.}~\bibnamefont {Liu}}, \bibinfo
  {author} {\bibfnamefont {X.-Y.}\ \bibnamefont {Yang}}, \bibinfo {author}
  {\bibfnamefont {W.-W.}\ \bibnamefont {Yu}}, \ and\ \bibinfo {author}
  {\bibfnamefont {Y.}~\bibnamefont {Zhang}},\ }\href {\doibase
  10.1103/PhysRevD.109.063526} {\bibfield  {journal} {\bibinfo  {journal}
  {Phys. Rev. D}\ }\textbf {\bibinfo {volume} {109}},\ \bibinfo {pages}
  {063526} (\bibinfo {year} {2024})},\ \Eprint
  {http://arxiv.org/abs/2308.15329} {arXiv:2308.15329 [astro-ph.CO]}
  \BibitemShut {NoStop}%
\bibitem [{\citenamefont {Nester}\ and\ \citenamefont
  {Yo}(1999)}]{Nester:1998mp}%
  \BibitemOpen
  \bibfield  {author} {\bibinfo {author} {\bibfnamefont {J.~M.}\ \bibnamefont
  {Nester}}\ and\ \bibinfo {author} {\bibfnamefont {H.-J.}\ \bibnamefont
  {Yo}},\ }\href@noop {} {\bibfield  {journal} {\bibinfo  {journal} {Chin. J.
  Phys.}\ }\textbf {\bibinfo {volume} {37}},\ \bibinfo {pages} {113} (\bibinfo
  {year} {1999})},\ \Eprint {http://arxiv.org/abs/gr-qc/9809049}
  {arXiv:gr-qc/9809049} \BibitemShut {NoStop}%
\bibitem [{\citenamefont {Beltr{\'a}n~Jim{\'e}nez}\ \emph
  {et~al.}(2019)\citenamefont {Beltr{\'a}n~Jim{\'e}nez}, \citenamefont
  {Heisenberg},\ and\ \citenamefont {Koivisto}}]{BeltranJimenez:2019esp}%
  \BibitemOpen
  \bibfield  {author} {\bibinfo {author} {\bibfnamefont {J.}~\bibnamefont
  {Beltr{\'a}n~Jim{\'e}nez}}, \bibinfo {author} {\bibfnamefont
  {L.}~\bibnamefont {Heisenberg}}, \ and\ \bibinfo {author} {\bibfnamefont
  {T.~S.}\ \bibnamefont {Koivisto}},\ }\href {\doibase 10.3390/universe5070173}
  {\bibfield  {journal} {\bibinfo  {journal} {Universe}\ }\textbf {\bibinfo
  {volume} {5}},\ \bibinfo {pages} {173} (\bibinfo {year} {2019})},\ \Eprint
  {http://arxiv.org/abs/1903.06830} {arXiv:1903.06830 [hep-th]} \BibitemShut
  {NoStop}%
\bibitem [{\citenamefont {Capozziello}\ \emph {et~al.}(2022)\citenamefont
  {Capozziello}, \citenamefont {De~Falco},\ and\ \citenamefont
  {Ferrara}}]{Capozziello:2022zzh}%
  \BibitemOpen
  \bibfield  {author} {\bibinfo {author} {\bibfnamefont {S.}~\bibnamefont
  {Capozziello}}, \bibinfo {author} {\bibfnamefont {V.}~\bibnamefont
  {De~Falco}}, \ and\ \bibinfo {author} {\bibfnamefont {C.}~\bibnamefont
  {Ferrara}},\ }\href {\doibase 10.1140/epjc/s10052-022-10823-x} {\bibfield
  {journal} {\bibinfo  {journal} {Eur. Phys. J. C}\ }\textbf {\bibinfo {volume}
  {82}},\ \bibinfo {pages} {865} (\bibinfo {year} {2022})},\ \Eprint
  {http://arxiv.org/abs/2208.03011} {arXiv:2208.03011 [gr-qc]} \BibitemShut
  {NoStop}%
\bibitem [{\citenamefont {Kobayashi}\ \emph {et~al.}(2016)\citenamefont
  {Kobayashi}, \citenamefont {Oikawa},\ and\ \citenamefont
  {Otsuka}}]{Kobayashi:2015aaa}%
  \BibitemOpen
  \bibfield  {author} {\bibinfo {author} {\bibfnamefont {T.}~\bibnamefont
  {Kobayashi}}, \bibinfo {author} {\bibfnamefont {A.}~\bibnamefont {Oikawa}}, \
  and\ \bibinfo {author} {\bibfnamefont {H.}~\bibnamefont {Otsuka}},\ }\href
  {\doibase 10.1103/PhysRevD.93.083508} {\bibfield  {journal} {\bibinfo
  {journal} {Phys. Rev. D}\ }\textbf {\bibinfo {volume} {93}},\ \bibinfo
  {pages} {083508} (\bibinfo {year} {2016})},\ \Eprint
  {http://arxiv.org/abs/1510.08768} {arXiv:1510.08768 [hep-ph]} \BibitemShut
  {NoStop}%
\bibitem [{\citenamefont {Cabo~Bizet}\ \emph {et~al.}(2016)\citenamefont
  {Cabo~Bizet}, \citenamefont {Loaiza-Brito},\ and\ \citenamefont
  {Zavala}}]{CaboBizet:2016uzv}%
  \BibitemOpen
  \bibfield  {author} {\bibinfo {author} {\bibfnamefont {N.}~\bibnamefont
  {Cabo~Bizet}}, \bibinfo {author} {\bibfnamefont {O.}~\bibnamefont
  {Loaiza-Brito}}, \ and\ \bibinfo {author} {\bibfnamefont {I.}~\bibnamefont
  {Zavala}},\ }\href {\doibase 10.1007/JHEP10(2016)082} {\bibfield  {journal}
  {\bibinfo  {journal} {JHEP}\ }\textbf {\bibinfo {volume} {10}},\ \bibinfo
  {pages} {082} (\bibinfo {year} {2016})},\ \Eprint
  {http://arxiv.org/abs/1605.03974} {arXiv:1605.03974 [hep-th]} \BibitemShut
  {NoStop}%
\bibitem [{\citenamefont {{\"O}zsoy}\ and\ \citenamefont
  {Lalak}(2021)}]{Ozsoy:2020kat}%
  \BibitemOpen
  \bibfield  {author} {\bibinfo {author} {\bibfnamefont {O.}~\bibnamefont
  {{\"O}zsoy}}\ and\ \bibinfo {author} {\bibfnamefont {Z.}~\bibnamefont
  {Lalak}},\ }\href {\doibase 10.1088/1475-7516/2021/01/040} {\bibfield
  {journal} {\bibinfo  {journal} {JCAP}\ }\textbf {\bibinfo {volume} {01}},\
  \bibinfo {pages} {040} (\bibinfo {year} {2021})},\ \Eprint
  {http://arxiv.org/abs/2008.07549} {arXiv:2008.07549 [astro-ph.CO]}
  \BibitemShut {NoStop}%
\bibitem [{\citenamefont {Akrami}\ \emph {et~al.}(2020)\citenamefont {Akrami}
  \emph {et~al.}}]{Planck:2018jri}%
  \BibitemOpen
  \bibfield  {author} {\bibinfo {author} {\bibfnamefont {Y.}~\bibnamefont
  {Akrami}} \emph {et~al.} (\bibinfo {collaboration} {Planck}),\ }\href
  {\doibase 10.1051/0004-6361/201833887} {\bibfield  {journal} {\bibinfo
  {journal} {Astron. Astrophys.}\ }\textbf {\bibinfo {volume} {641}},\ \bibinfo
  {pages} {A10} (\bibinfo {year} {2020})},\ \Eprint
  {http://arxiv.org/abs/1807.06211} {arXiv:1807.06211 [astro-ph.CO]}
  \BibitemShut {NoStop}%
\bibitem [{\citenamefont {Chen}\ \emph {et~al.}(2025)\citenamefont {Chen},
  \citenamefont {Liu},\ and\ \citenamefont {Zhang}}]{Chen:2024ikn}%
  \BibitemOpen
  \bibfield  {author} {\bibinfo {author} {\bibfnamefont {J.}~\bibnamefont
  {Chen}}, \bibinfo {author} {\bibfnamefont {C.}~\bibnamefont {Liu}}, \ and\
  \bibinfo {author} {\bibfnamefont {Y.-L.}\ \bibnamefont {Zhang}},\ }\href
  {\doibase 10.1088/1475-7516/2025/05/050} {\bibfield  {journal} {\bibinfo
  {journal} {JCAP}\ }\textbf {\bibinfo {volume} {05}},\ \bibinfo {pages} {050}
  (\bibinfo {year} {2025})},\ \Eprint {http://arxiv.org/abs/2410.18916}
  {arXiv:2410.18916 [gr-qc]} \BibitemShut {NoStop}%
\bibitem [{\citenamefont {Calabrese}\ \emph {et~al.}(2025)\citenamefont
  {Calabrese} \emph {et~al.}}]{ACT:2025tim}%
  \BibitemOpen
  \bibfield  {author} {\bibinfo {author} {\bibfnamefont {E.}~\bibnamefont
  {Calabrese}} \emph {et~al.} (\bibinfo {collaboration} {ACT}),\ }\href@noop {}
  {\  (\bibinfo {year} {2025})},\ \Eprint {http://arxiv.org/abs/2503.14454}
  {arXiv:2503.14454 [astro-ph.CO]} \BibitemShut {NoStop}%
\bibitem [{\citenamefont {Vallisneri}(2008)}]{Vallisneri:2007ev}%
  \BibitemOpen
  \bibfield  {author} {\bibinfo {author} {\bibfnamefont {M.}~\bibnamefont
  {Vallisneri}},\ }\href {\doibase 10.1103/PhysRevD.77.042001} {\bibfield
  {journal} {\bibinfo  {journal} {Phys. Rev. D}\ }\textbf {\bibinfo {volume}
  {77}},\ \bibinfo {pages} {042001} (\bibinfo {year} {2008})},\ \Eprint
  {http://arxiv.org/abs/gr-qc/0703086} {arXiv:gr-qc/0703086} \BibitemShut
  {NoStop}%
\bibitem [{\citenamefont {Su}\ \emph {et~al.}(2025)\citenamefont {Su},
  \citenamefont {Xu}, \citenamefont {Chen}, \citenamefont {Liu},\ and\
  \citenamefont {Zhang}}]{Su:2025nkl}%
  \BibitemOpen
  \bibfield  {author} {\bibinfo {author} {\bibfnamefont {H.}~\bibnamefont
  {Su}}, \bibinfo {author} {\bibfnamefont {B.}~\bibnamefont {Xu}}, \bibinfo
  {author} {\bibfnamefont {J.}~\bibnamefont {Chen}}, \bibinfo {author}
  {\bibfnamefont {C.}~\bibnamefont {Liu}}, \ and\ \bibinfo {author}
  {\bibfnamefont {Y.-L.}\ \bibnamefont {Zhang}},\ }\href {\doibase
  10.1088/1572-9494/add1b9} {\bibfield  {journal} {\bibinfo  {journal} {Commun.
  Theor. Phys.}\ }\textbf {\bibinfo {volume} {77}},\ \bibinfo {pages} {115403}
  (\bibinfo {year} {2025})},\ \Eprint {http://arxiv.org/abs/2503.20778}
  {arXiv:2503.20778 [astro-ph.CO]} \BibitemShut {NoStop}%
\end{thebibliography}%

%%%%%%%%%%%%%%%%%%%%%%%%%%%%%%%%%%%%%%%%%%%%%%%%%%%%%%%%%%%%%%%%%%%%%%%%%%%%%%%%
\end{document}